\documentstyle{mn}
 
\newif\ifAMStwofonts
 
\newcommand{\beq}{\begin{equation}}
\newcommand{\eeq}{\end{equation}}
 
\newcommand{\figl}[7]{
    \protect\centerline{
    \epsfxsize=#1\epsffile[#2 #3 #4 #5]{#6 angle=#7}}}
 
\def\lsim{\mathrel{\lower2.5pt\vbox{\lineskip=0pt\baselineskip=0pt
           \hbox{$<$}\hbox{$\sim$}}}}
\def\gsim{\mathrel{\lower2.5pt\vbox{\lineskip=0pt\baselineskip=0pt
           \hbox{$>$}\hbox{$\sim$}}}}

\ifoldfss
    \NewTextAlphabet{textbfit} {cmbxti10} {}
    \NewTextAlphabet{textbfss} {cmssbx10} {}
    \NewMathAlphabet{mathbfit} {cmbxti10} {} 
    \NewMathAlphabet{mathbfss} {cmssbx10} {} 
  \fi
  \ifAMStwofonts
    \ifCUPmtlplainloaded \else
      \NewSymbolFont{upmath} {eurm10}
      \NewSymbolFont{AMSa} {msam10}
      \NewMathSymbol{\upi}     {0}{upmath}{19}
      \NewMathSymbol{\umu}     {0}{upmath}{16}
      \NewMathSymbol{\upartial}{0}{upmath}{40}
      \NewMathSymbol{\leqslant}{3}{AMSa}{36}
      \NewMathSymbol{\geqslant}{3}{AMSa}{3E}

       \let\le=\leqslant
      \let\geq=\geqslant \let\ge=\geqslant
    \fi
  \fi

\ifnfssone
  \newmathalphabet{\mathit}
  \addtoversion{normal}{\mathit}{cmr}{m}{it}
  \addtoversion{bold}{\mathit}{cmr}{bx}{it}
  \newmathalphabet{\mathbfit} 
  \addtoversion{normal}{\mathbfit}{cmr}{bx}{it}
  \addtoversion{bold}{\mathbfit}{cmr}{bx}{it}
  \newmathalphabet{\mathbfss} 
  \addtoversion{normal}{\mathbfss}{cmss}{bx}{n}
  \addtoversion{bold}{\mathbfss}{cmss}{bx}{n}
  \ifAMStwofonts
    \ifCUPmtlplainloaded \else
      \UseAMStwoboldmath
      \makeatletter
      \new@mathgroup\upmath@group
      \define@mathgroup\mv@normal\upmath@group{eur}{m}{n}
      \define@mathgroup\mv@bold\upmath@group{eur}{b}{n}
      \edef\UPM{\hexnumber\upmath@group}
      \new@mathgroup\amsa@group
      \define@mathgroup\mv@normal\amsa@group{msa}{m}{n}
      \define@mathgroup\mv@bold\amsa@group{msa}{m}{n}
      \edef\AMSa{\hexnumber\amsa@group}
      \makeatother
      \mathchardef\upi="0\UPM19
      \mathchardef\umu="0\UPM16
      \mathchardef\upartial="0\UPM40
      \mathchardef\leqslant="3\AMSa36
      \mathchardef\geqslant="3\AMSa3E

       \let\le=\leqslant
      \let\geq=\geqslant \let\ge=\geqslant
    \fi
  \fi
\fi 

\ifnfsstwo
  \DeclareMathAlphabet{\mathbfit}{OT1}{cmr}{bx}{it}
  \SetMathAlphabet\mathbfit{bold}{OT1}{cmr}{bx}{it}
  \DeclareMathAlphabet{\mathbfss}{OT1}{cmss}{bx}{n}
  \SetMathAlphabet\mathbfss{bold}{OT1}{cmss}{bx}{n}
  \ifAMStwofonts
    \ifCUPmtlplainloaded \else
      \DeclareSymbolFont{UPM}{U}{eur}{m}{n}
      \SetSymbolFont{UPM}{bold}{U}{eur}{b}{n}
      \DeclareSymbolFont{AMSa}{U}{msa}{m}{n}
      \DeclareMathSymbol{\upi}{0}{UPM}{"19}
      \DeclareMathSymbol{\umu}{0}{UPM}{"16}
      \DeclareMathSymbol{\upartial}{0}{UPM}{"40}
      \DeclareMathSymbol{\leqslant}{3}{AMSa}{"36}
      \DeclareMathSymbol{\geqslant}{3}{AMSa}{"3E}

       \let\le=\leqslant
      \let\geq=\geqslant \let\ge=\geqslant
    \fi
  \fi
\fi 

\ifCUPmtlplainloaded \else
  \ifAMStwofonts \else 
    \def\upi{\pi}
    \def\umu{\mu}
    \def\upartial{\partial}
  \fi
\fi

\input epsf

\title[Effectiveness of sub-mm source 
redshift estimation based on rest-frame radio--FIR photometry]{
Confirmation of the effectiveness of sub-mm source redshift 
estimation based on rest-frame radio--FIR photometry}
\author[Aretxaga, Hughes, Dunlop]
{Itziar Aretxaga$^{1}$,  David H. Hughes$^{1}$, James S. Dunlop$^{2}$\\
$^{1}$Instituto Nacional de Astrof\'{\i}sica, \'Optica y Electr\'onica 
(INAOE), Aptdo. Postal 51 y 216, 72000 Puebla, Mexico \\
$^{2}$Institute for Astronomy, University of Edinburgh, Royal Observatory, 
Edinburgh, EH9 3HJ, UK \\
}

\pagerange{\pageref{firstpage}--\pageref{lastpage}}
\pubyear{2004}

\begin{document}

\maketitle 

\label{firstpage}

\begin{abstract}
We present a comparison between the published optical, IR and CO
spectroscopic redshifts of 15 (sub-)mm galaxies and their photometric
redshifts as derived from long-wavelength (radio--mm--FIR) photometric
data. The redshift accuracy measured for 12 sub-mm galaxies with at least one
robustly-determined colour in the radio--mm--FIR regime is $\delta z
\approx 0.30$ (r.m.s.).  Despite the wide range of spectral energy
distributions in the local galaxies that are used in an un-biased
manner as templates, this analysis demonstrates that photometric redshifts
can be efficiently derived for sub-mm galaxies with a precision of
$\delta z < 0.5$ using only the rest-frame FIR to radio wavelength
data.

\end{abstract}


\section{Introduction}

The next generation of wide-area extragalactic submillimetre and
millimetre (hereafter sub-mm) surveys, for example from the
Balloon-borne Large Aperture Submillimetre Telescope (BLAST, Devlin et
al 2001), LABOCA on the Atacama Pathfinder Experiment
(APEX\footnote{www.mpifr-bonn.mpg.de/div/mm/apex/}), the SCUBA\,2
camera\footnote{www.roe.ac.uk/ukatc/projects/scubatwo/index.html} on
the James Clerk Maxwell Telescope (JCMT) and BOLOCAM-II on the Large
Millimetre Telescope (LMT\footnote{www.lmtgtm.org}), will
produce large samples ($\sim 10^{3} - 10^{5}$) of distant, luminous
starburst galaxies.  The dramatic increase in the number of
submillimetre detected galaxies requiring follow-up observations makes
it unreasonable to expect that a large fraction of their obscured or
faint optical and IR counterparts will have unambiguous,
spectroscopically-determined redshifts. An alternative method to
efficiently and robustly measure the redshift distribution for large
samples of submillimetre galaxies is clearly
necessary.

Given the underlying assumption that we are witnessing high rates of
star formation in these submillimetre galaxies, then we expect them to
have the characteristic FIR peak and steep submillimetre
(Rayleigh-Jeans) spectrum which is dominated by thermal emission from
dust heated to temperatures in the range $\sim 20-70$\,K by obscured
young, massive stars.  The observed radio--FIR luminosity correlation 
in local starburst galaxies (e.g. Helou et al. 1985), that
links the radio synchrotron emission from supernova remnants with the
later stages of massive star formation, is also expected to apply to
the submillimetre galaxies.

Thus, in recent years, a considerable amount of effort has been
invested in assessing the accuracy with which these broad continuum
features in the spectral energy distributions (SEDs) of submillimetre
galaxies at rest-frame mid-IR to radio wavelengths can be used to
provide photometric-redshifts (Hughes et al. 1998; Carilli \& Yun
1999, 2000; Dunne, Clements \& Eales 2000; Rengarajan \& Takeuchi
2001; Yun \& Carilli 2002; Wiklind 2003; Blain, Barnard \& Chapman
2003).  In a contribution to this general investigation Hughes et
al. (2002, Paper\,I) described Monte-Carlo simulations that used a
library of multi-frequency template SEDs, derived from observations of
local starbursts and AGN with a wide-range of FIR luminosities ($9.0 <
{\rm log} L_{\rm FIR}/L_{\odot} < 12.3$) and temperatures
($25<T/K<65$), to measure the accuracy of photometric redshifts that
could be derived from future 250, 350, 500$\mu$m extragalactic surveys
with BLAST and Herschel, and complementary 850$\mu$m surveys from
SCUBA.  Aretxaga et al. (2003, Paper\,II) then applied the techniques
described in Paper\,I to the catalogues of submillimetre galaxies
identified in various SCUBA surveys, and derived photometric redshifts
for individual sources using existing radio--submillimetre data.

In this paper we use new optical and IR spectroscopic observations of  
submillimetre galaxies published by Chapman et al. (2003a), and
Simpson et al. (2004) to update our previous comparison of spectroscopic 
and long-wavelength photometric redshifts (Paper II).
In section 2 we explain the selection criteria 
for the sub-mm sources with 
spectroscopic redshifts 
included in our analysis, and 
demonstrate how the accuracy of the photometric redshift prediction varies 
according to the quality of the radio--mm--FIR photometric data.
In section 3 we discuss the significant agreement between the
spectroscopic and photometric redshifts, and use these results to
challenge the suggestion by Blain et al. (2003) that it is not
possible to derive photometric redshifts with an accuracy of $dz
\simeq \pm 0.5$ or better without adopting an unreasonably tight
dispersion in the luminosity-temperature (L-T) relation or a limited
range of SEDs in the analysis.
Finally, our conclusions are summarized in section\,4.

The cosmological parameters adopted throughout this paper are 
$H_0=67$~km\,s$^{-1}$\,Mpc$^{-1}$, $\Omega_{\rm M}=0.3$,
$\Omega_{\Lambda}=0.7$.

\section{Analysis}

\subsection{New spectroscopic redshifts for sub-mm sources}

Over the last year several new optical and IR spectroscopic redshifts
have been published for sub-mm sources (Chapman et al. 2003a, Simpson
et al. 2004), enhancing substantially the number of sources available
for checking the reliability of photometric estimates.  Many of the
sub-mm galaxies selected for spectroscopic study have been extracted
from the 8-mJy SCUBA survey (Scott et al. 2002), aided by deep
follow-up observations with the VLA (Ivison et al. 2002).

Of the 10 optical redshifts published by Chapman et al. (2003a), we
include in our analysis only those 5 sources with photometric data of
sufficient quality. We briefly discuss the SED properties of the
excluded sources:

\begin{itemize}

\item SMM105224.6+572119 ($z_{opt}=2.429$), also known as LE850.15
from the UK 8-mJy SCUBA survey of the Lockman Hole East, is rejected
because the sub-mm source has since been shown to be a spurious
850$\mu$m detection.  In Scott et al. (2002) this source was
originally listed as a low S/N (3.4$\sigma$) submm source, having been
extracted from one of the more noisy regions of the 8-mJy
maps. Subsequently it was highlighted by Ivison et al. (2002) to be
very likely erroneous on account of its lack of a radio counterpart
(despite its very bright apparent sub-mm flux).  The spurious nature
of this source has been supported by the failure to detect the source
in a reanalysis of the 8-mJy maps conducted as part of SHADES (Mortier
et al. 2004; {\tt www.roe.ac.uk/$\sim$ifa/shades}), and by its
non-detection in the IRAM MAMBO mm-wavelength maps of the 8-mJy
Lockman East field (Greve et al. 2004).

\item SMM105207.7+571907 ($z_{opt}=2.698$) or LE850.12, also from the UK 8-mJy
SCUBA survey, has a bright and statistically secure radio counterpart
(Ivison et al. 2002).
This object, however, is a known AGN, with strong, flat-spectrum,
variable radio emission. Therefore, it must be excluded from the analysis
presented here, which relies on the fitting of
non-variable 
starburst-dominated spectral templates.  
As we show in Fig~3, no SED in our local template library can
reproduce the high radio fluxes of this variable radio-source at the
spectroscopic redshift, and indeed, in Paper II we already reported
that the SED could not be matched at any redshift with the local
templates.

\item SMM123600.2+621047 ($z_{opt}=1.998$), whose position (J2000)
corresponds to a 1.4\,GHz radio source (123600.150+621047.17) in
Richards (2000), has a spectrum which presumably arises from an
optically-faint ($I=23.6$) VLA radio-source that was followed up with
submillimetre photometry, but the robustness of the radio-submm
association cannot be 
determined from the published information.

\item \mbox{SMM131201.2+424208 ($z_{opt}=3.419$) and} SMM131212.7+424423 ($z_{opt}=2.811$) still do not have published 1.4GHz data
from the SSA13 field (Fomalont et al., in prep), and we therefore
cannot reconstruct the radio--sub-mm SED of this source.

\end{itemize}

In the analysis described in \S\,2.2 we include the remaining 5 
sub-mm sources from the 8-mJy survey studied by Chapman et
al. (2003a), namely N2850.1, N2850.4, N2850.8, LE850.6 and LE850.18,
which have unambiguous radio identifications from Ivison et
al. (2002). The spectroscopic redshifts for all these sources also
appear to be secure, and indeed redshift information for two of them
has been published elsewhere: N2850.4 (Smail et al. 2003)
and N2850.1 (Chapman et al. 2002a)
 
From deep near-infrared spectroscopy with the Subaru telescope,
Simpson et al. (2004) have measured spectroscopic redshifts based on
two or more emission lines with secure radio and near-infrared
identifications for 2 additional 8-mJy SCUBA sources (N2850.2 and
N2850.12 in the ELAIS N2 Field, Scott et al. 2002).  We therefore
include these two sources in the expanded sample analysed here.  We
note that Simpson et al. also suggested a tentative redshift for
LE850.3 estimated from a single 2.5$\sigma$ emission-line, and also a
redshift for CUDSS14.9 from the putative HK Ca absorption lines that
need confirmation (Simpson, priv. communication). These latter two
redshifts are not considered robust enough for inclusion in our
analysis.

\subsection{New photometric redshifts} 

We thus have 7 new redshifts for robust sub-mm sources
from the UK 8-mJy SCUBA survey, which can be added to the heterogenous 
collection of 8 sub-mm sources
with spectroscopic redshifts previously considered by Aretxaga et al.
(2003). Thus, the new spectroscopic data have effectively doubled the size of 
the comparison sample, as well as extending the redshift baseline out 
to $z \simeq 4$.

\begin{table*}
 \begin{minipage}{170mm}
\begin{center}
\caption{Photometric redshifts for the sources with reported
spectroscopic redshifts.  The first column gives the name; the second
column gives the most probable mode and the 68\% confidence interval
for the mode among 100 Monte Carlo realizations; the third column
gives the 68\% confidence interval for the redshift distribution of
the source; the fourth one, the 90\% confidence interval; the fifth
and sixth columns give respectively the detected bands (at a $\ge
3\sigma$ level) and the upper-limits used for the computation of the
photometric-redshifts; and the seventh column gives the spectroscopic
redshifts and their references: Ba99 for Barger et
al. 1999; Ch02a for Chapman et al. 2002a; Ch02b for Chapman et
al. 2002b; Ch02c for Chapman et al. 2002c; Ch03 for Chapman et
al. 2003a; Ea00 for Eales et al. 2000; Fr98 for Frayer et al. 1998;
HR94 for Hu \& Ridgway 1994; Si04 for Simpson et al. 2004; and Sm03
for Smail et al. 2003.  }
\label{3filters}
\begin{tabular}{lcccccl}
\hline
  object & $z_{\rm phot}$ mode & 68\% CL & 90\% CL & $\geq
  3\sigma$ detections & $< 3\sigma$ / upper limits & $z_{\rm
  spec}$ \\
\hline

LH850.6 & 2.6$\pm^{0.4}_{0.0}$ & (2.5--3.7) & (2.0--4.2) &
850$\mu$m,1.2mm,1.4GHz & 450$\mu$m,5GHz & 2.610 (Ch03)\\ 

LH850.18 & 3.8$\pm^{0.2}_{0.3}$ & (3.0--5.4) & (2.6--6.0) & 
850$\mu$m,1.2mm,1.4GHz & 450$\mu$m,5GHz & 
3.699 (Ch03)\\

N2850.1 & 2.85$\pm^{0.05}_{0.15}$ & (2.5--3.8) & (2.0--4.1) & 450,850$\mu$m,1.4GHz & 1.2mm & 0.840 (Ch02a)
\\ 

N2850.2 & 2.35$\pm^{0.05}_{0.15}$ & (2.0--2.7) & (1.5--3.1) & 450, 850$\mu$m,1.2mm,1.4GHz & --- & 2.45 (Si04) \\

N2850.4 & 2.6$\pm^{0.1}_{0.2}$ & (2.0--3.8) & (2.0--4.6) & 850$\mu$m,1.2mm,1.4GHz & 450$\mu$m & 2.376 (Sm03)
\\

N2850.8  &        2.5$\pm0.3$   & (1.5--3.5) & (1.0--4.5) & 850$\mu$m & 450$\mu$m,1.2mm,1.4GHz & 1.189 (Ch03)\\

N2850.12 &        2.5$\pm^{0.2}_{0.1}$  & (1.6--4.0) & (1.5--5.0) &  850$\mu$m & 450$\mu$m,1.2mm,1.4GHz & 2.43 (Si04)\\


CUDSS14.18  & 0.7$\pm^{1.1}_{0.5}$ & 
(0.3--1.5) & (0.0--1.8) & 5,1.4GHz & 450,850$\mu$m & 0.66 (Ea00)\\

SMMJ02399$-$0134 &2.3$\pm^{0.3}_{1.8}$ & (1.5--3.1) & (1.0--3.8) & 
$850\mu$m,1.4GHz & --- & 1.056 (Ba99)  \\

SMMJ02399$-$0136 & 2.85$\pm 0.05$ & (2.5--3.7) & (2.0--5.0) &
450,850$\mu$m,1.3mm,1.4GHz & 2mm & 2.808 (Fr98) \\

SMMJ14011$+$0252 & 2.85$\pm^{0.05}_{0.35}$ & (2.0--3.3) & (2.0--4.4) & 
450,850$\mu$m,1.3mm,1.4GHz & 3mm,2.8GHz & 2.550 (Ba99)\\

W$-$MM11 & 3.7$\pm1.0$& (2.5--5.3) & (2.0--5.8) & 
850$\mu$m& 450$\mu$m & 2.98 (Ch02b) \\

HR10 & 1.95$\pm^{0.05}_{0.25}$ & (1.5--2.3) & (1.5--2.9) & 
450,850$\mu$m,1.3mm,8GHz & 100$\mu$m,1.4GHz & 1.44 (HR94) \\

N1$-40$ & 0.8$\pm^{0.2}_{0.3}$ & (0.5--1.2) & (0.5--1.5) &
175,850$\mu$m,1.4GHz & 450$\mu$m & 0.45 (Ch02c) \\

N1$-64$ & 1.05$\pm^{0.35}_{0.25}$& (0.9--2.0)& (0.5--2.1)& 
175,850$\mu$m,1.4GHz & 450$\mu$m & 0.91 (Ch02c) \\
\hline
\end{tabular}
\end{center}
\end{minipage}
\end{table*}

\begin{figure}

\hspace*{7cm} 
\figl{7cm}{89}{192}{550}{655}{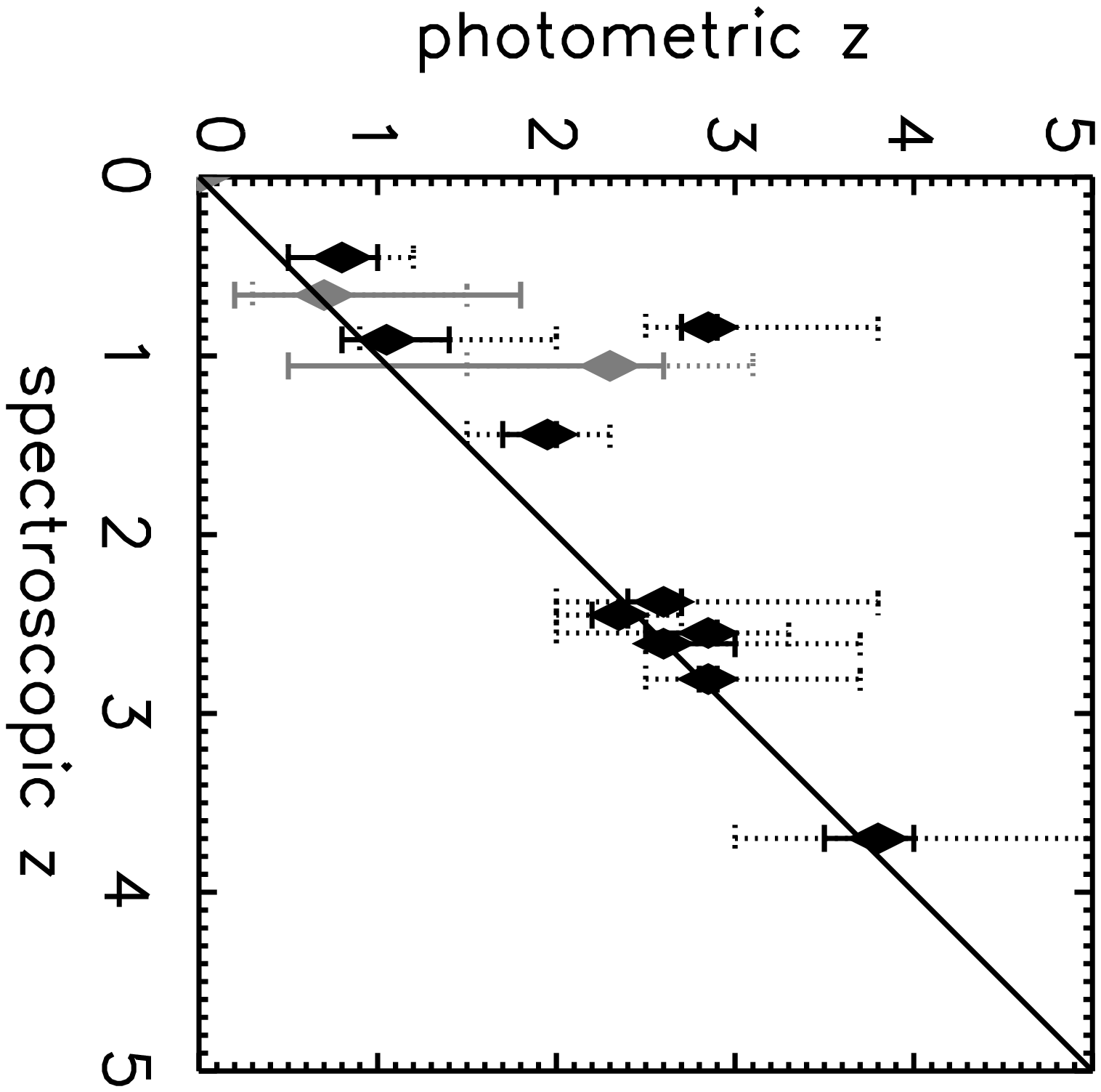}{90}\\
\vspace*{0.5cm}
\caption{Comparison of the new photometric-redshifts, using model 
le2 as a prior (see Paper~II) and the true redshifts for the 12 sub-mm
sources with at least one robust radio/sub-mm/far-infrared colour.
The diamonds represent the most probable mode derived from 100 Monte
Carlos calculated for each object using model le2. The solid error
bars represent the range of modes (68\% confidence level) measured in the
corresponding redshift distributions of the 100 realizations, and the
dotted lines show the 68\% confidence interval of a representative
redshift distribution for each object.  Sources represented in black
(in increasing redshift: N1--40, N2850.1, N1--64, HR10, N2850.4,
N2850.2, SMMJ14011+0252 , LE850.6, SMMJ02399$-$0136 and LE850.18),
have photometric redshifts derived from measurements ($\ge 3\sigma$)
in at least three passbands of the radio--mm--FIR regime with the
addition of some upper limits, and are the most precise estimates.  Sources
represented in dark grey (in increasing redshift: CUDSS14.18,
SMMJ02399--0134) have photometric redshifts derived
from measurements ($\ge 3\sigma$) in just two passbands and some
additional upper limits. }
\label{photz_z}
\end{figure}

Since our original analysis in Paper~II, new photometric data at 1.2mm
obtained with MAMBO (Greve et al. 2004) have also become available for the
sub-mm galaxies first identified in the UK 8-mJy SCUBA survey areas
(Scott et al. 2002).  We have therefore incorporated this additional
photometric information in a re-derivation of the photometric
redshifts of UK 8-mJy SCUBA sources in the current spectroscopic
sample.  
We also include N2850.12 which has a
3.4$\sigma$ flux determination at 850$\mu$m, and hence failed our
selection criteria in Paper~II, but which now has a
spectroscopic-redshift (Simpson et al. 2004) based on an IR
counterpart associated with the 3$\sigma$ peak in the radio
observations of Ivison et al. (2002).

We have adopted the fiducial evolutionary model le2 (the least
restrictive one in Paper~II, which applied a non-informative prior at
$z>2.3$, and $(1+z)^3$ luminosity evolution to $z\le 2.3$) to derive
the new photometric-redshift estimates which are given in Table~1.
We produce simulated catalogs of sources in mock surveys to the same
depth as those in which the real sub-mm galaxies were detected. The
corresponding redshift distributions are then derived by the joint
probability of identifying the particular radio--mm--FIR 
colours and fluxes of the 
real sources with the mock galaxies in the simulated catalogs.
  In order to check the stability of the
redshift solutions found, we have calculated 100 different Monte
Carlos for each source, and for each of these realizations we have
derived its redshift probability distribution, its mode and 68\% and
90\% confidence level intervals.  As expected, the sources with
photometric redshifts based on 2 or more 
colours are the most stable
and show well defined peaks and little variation in the mode of the redshift
distribution of the different
simulations.  
It is important to note  that these photometric redshift
estimates include absolute calibration errors of 5-20\% in the
individual radio--submillimetre fluxes, and assume no correlation between the
luminosity and shape of an SED (or temperature).

From our sample of 15 sub-mm galaxies, Fig.~1 shows the
photometric-redshift spectroscopic-redshift regression plot for those
12 sources that have at least one robust colour based on two or more
detections ($\ge 3\sigma$) in the radio-mm-FIR regime.
The remaining 3 sources (N2850.8,
N2850.12, and W-MM11) have a robust detection at only one wavelength
and various upper-limits and hence are not included in Fig.~1.

Figs.~2 and 3 show the comparison of the observed SEDs of the sources
analysed in this paper and the template SEDs considered in the
photometric redshift analysis, illustrating the cases where there is good
agreement and catastrophic disagreement, respectively.  All SEDs in
our template library are accepted by more than one of these sources.

\subsubsection{The case of N2850.1}

Among the photometric redshifts based on at least two colours (with
well measured fluxes, i.e. $\ge 3\sigma$, in three or more bands) the
submillimetre galaxy which departs most clearly from the $z_{\rm
phot}=z_{\rm spec}$ line is N2850.1 ($z_{phot}-z_{spec}=2.01$).  Our
new photometric redshift of N2850.1, using the most recent upper limit
at 1.2mm (Greve et al. 2004) together with the 450, 850$\mu$m and
1.4GHz fluxes, is $z_{phot} = 2.8\pm^{1.0}_{0.3}$ at a 68\% confidence
level, and $z_{phot} = 2.8\pm^{1.0}_{0.8}$ at a 90\% level.  These
values are consistent with the estimates presented in Paper~II, but
remain strongly inconsistent with the optical spectroscopic
redshift ($z_{\rm spec}=0.840$) reported by (Chapman et al. 2003a).
This is not surprising as there has already been considerable debate
over whether the optical identification for this sub-mm source
is correct.  The optical spectrum and redshift 
for this
counterpart, originally published by Chapman et al. (2002a), led these 
authors to
argue that the optically-bright galaxy coincident (within $\sim
0.2''$) with the radio position of N2850.1 was possibly a foreground
galaxy that lenses the sub-mm source. This argument was based primarily on the
fact that the temperature of the dust emission $T_{\rm D}=23\pm 5$~K
deduced for the sub-mm source at the optical spectroscopic redshift
was 4$\sigma$ below that of the local dusty galaxies with the same
intrinsic luminosity.  In Chapman et al. (2003a), N2850.1 has a
revised temperature of $16^{+4.1}_{-2.2}$~K or 6$\sigma$ below the
temperature distribution of local analogs, and colder than
SMM22173+0014, which Chapman et al. (2002a) also claim is lensed.

It is clear therefore that among the sources considered here, N2850.1 
is the most likely example of a bright sub-mm source produced by gravitational
lensing by an intermediate redshift galaxy, analogous to the case of HDF850.1 
studied in detail by Dunlop et al. (2004). Proving this beyond doubt remains a 
challenge, as astrometric information of the quality available to Dunlop
et al. (2004) does not yet exist in the case of N2850.1.

Whatever the correct explanation for its properties, it is obvious that
N2850.1 is an unusual source, being the only object in the sample 
of 15 galaxies which cannot be fit by any of our SED templates
when redshifted to $z=0.840$ (see Fig.~3). For
the sake of transparency and completeness in what follows we therefore
quote the overall accuracy of the redshift estimation procedure both
with and without the inclusion of N2850.1 in the statistical
calculations.

\subsubsection{Overall accuracy of photometric redshifts}

In general the agreement between photometric and spectroscopic
redshifts is encouraging.  The three sources for which
photometric-redshifts are based on a solid detection ($\ge 3\sigma$)
at just one wavelength, however,
are the least precise due to insufficient photometric constraints, although
we note that they are still formally
consistent with the spectroscopic redshifts found: N2850.8 at $z_{\rm
spec}=1.189$ has a photometric redshift $z_{\rm phot}=2.5\pm1.0$ at a
68\% confidence level (1.0--4.5 at the 90\% level); N2850.12 has a
photometric redshift of $z_{\rm phot}=2.5\pm^{1.5}_{0.9}$ at a 68\%
confidence level (1.5 to 5.0 at the 90\% level), which is consistent
with the $z_{\rm spec}=2.43$ (Simpson et al. 2004); and W-MM11 at
$z_{\rm spec}=2.98$, as determined in Paper~II, has a $z_{\rm
phot}=3.7\pm^{1.6}_{1.2}$ at a 68\% confidence level (2.0--5.8 at the
90\% level).

Despite the strong suspicion that N2850.1 is a lensed SCUBA galaxy at
a redshift $z \gg z_{opt} = 0.840$, even if we include N2850.1 in our
analysis the rms dispersion about $z_{phot}=z_{spec}$ is $\delta z=
0.38$.  This result considers only those 10/15 sub-mm galaxies with at
least two measured colours based on 3 or more $\ge 3\sigma$ detections
in the radio--mm--FIR regime.  The precision significantly improves to
$\delta z= 0.20$ if we exclude N2850.1. Extending this analysis to the
12/15 sources with at least one colour determined from detections at
two or more bands, the measured mean accuracy of the photometric
redshifts in the range $0.5 < z < 4$ is $\delta z= 0.42$ and $\delta
z= 0.28$, including and excluding N2850.1, respectively. Finally if we
also include sources with only upper-limits in the colours (for
example a single submillimetre detection with a non-detection at radio
wavelengths), we measure $\delta z= 0.48$ and $\delta z= 0.37$,
including and excluding N2850.1, respectively.

\begin{figure*}
    \vspace*{-1.0cm}
    \hbox{ 
           \hspace{2.5cm} \figl{4.5cm}{26}{26}{573}{737}{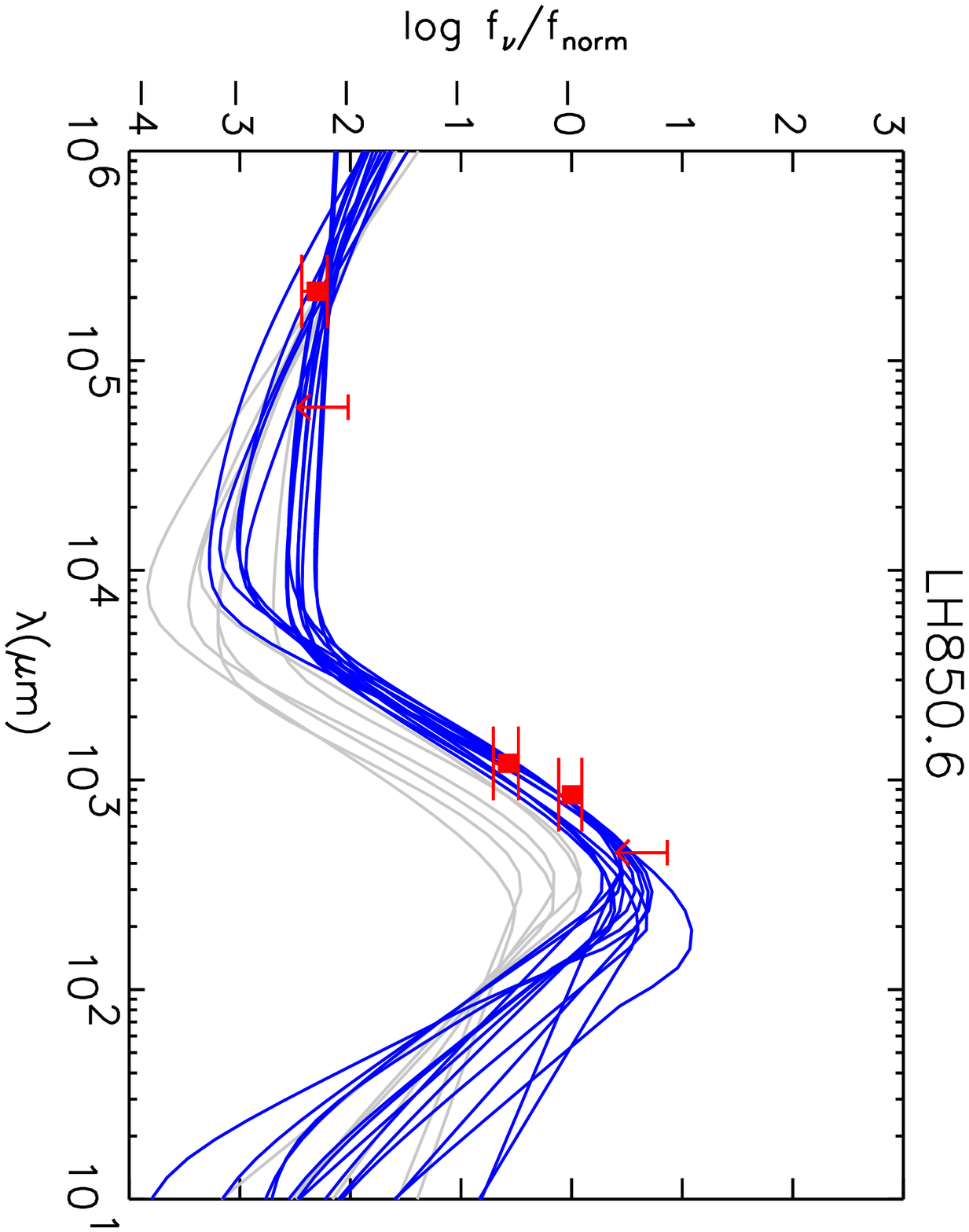}{90}
           \hspace{-13cm}  \figl{4.5cm}{26}{26}{573}{737}{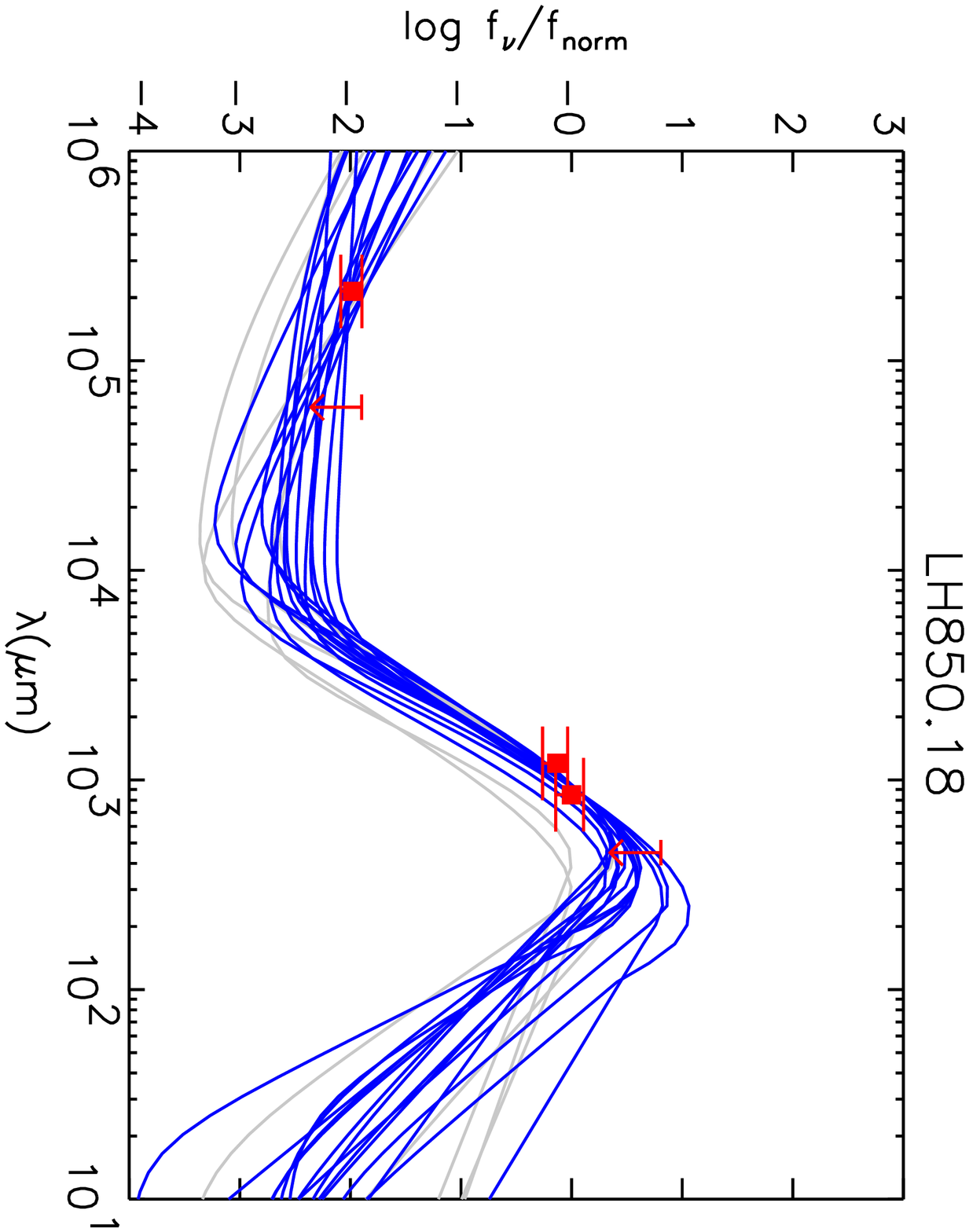}{90} 
           }

    \vspace*{-1.7cm}
    \hbox{ \hspace{0.cm} \figl{4.5cm}{26}{26}{573}{737}{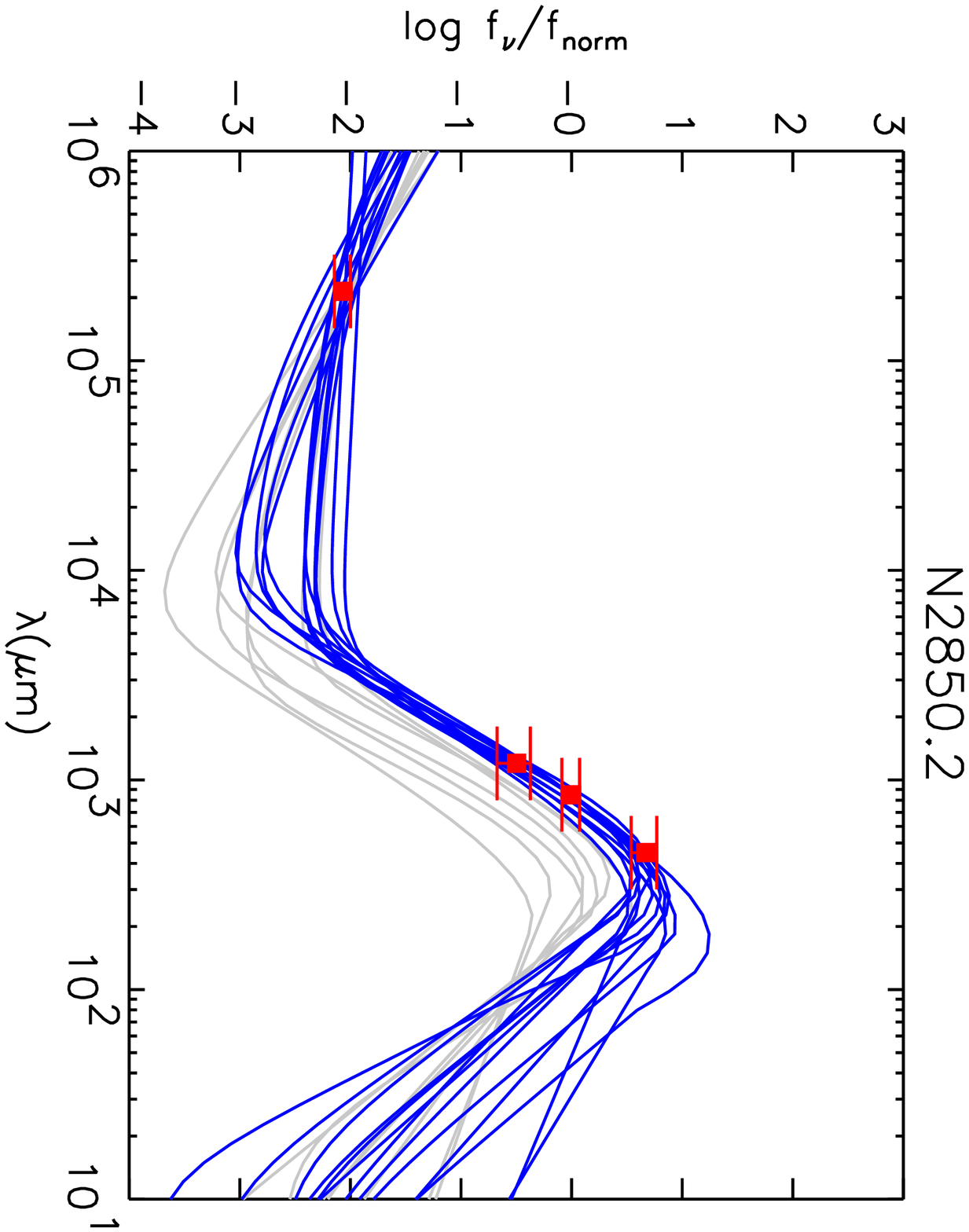}{90}
           \hspace{-13cm} \figl{4.5cm}{26}{26}{573}{737}{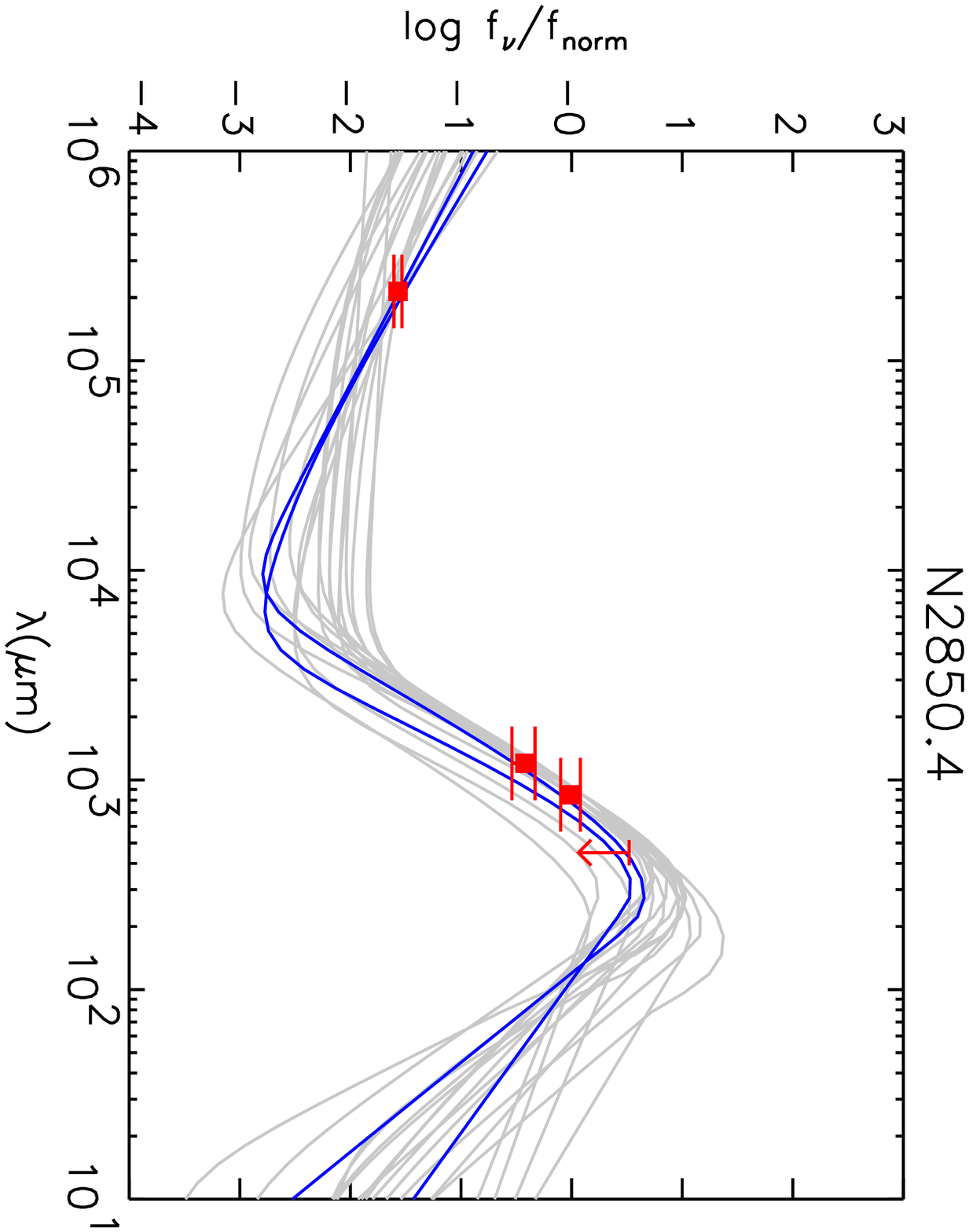}{90}
          \hspace{-13cm} \figl{4.5cm}{26}{26}{573}{737}{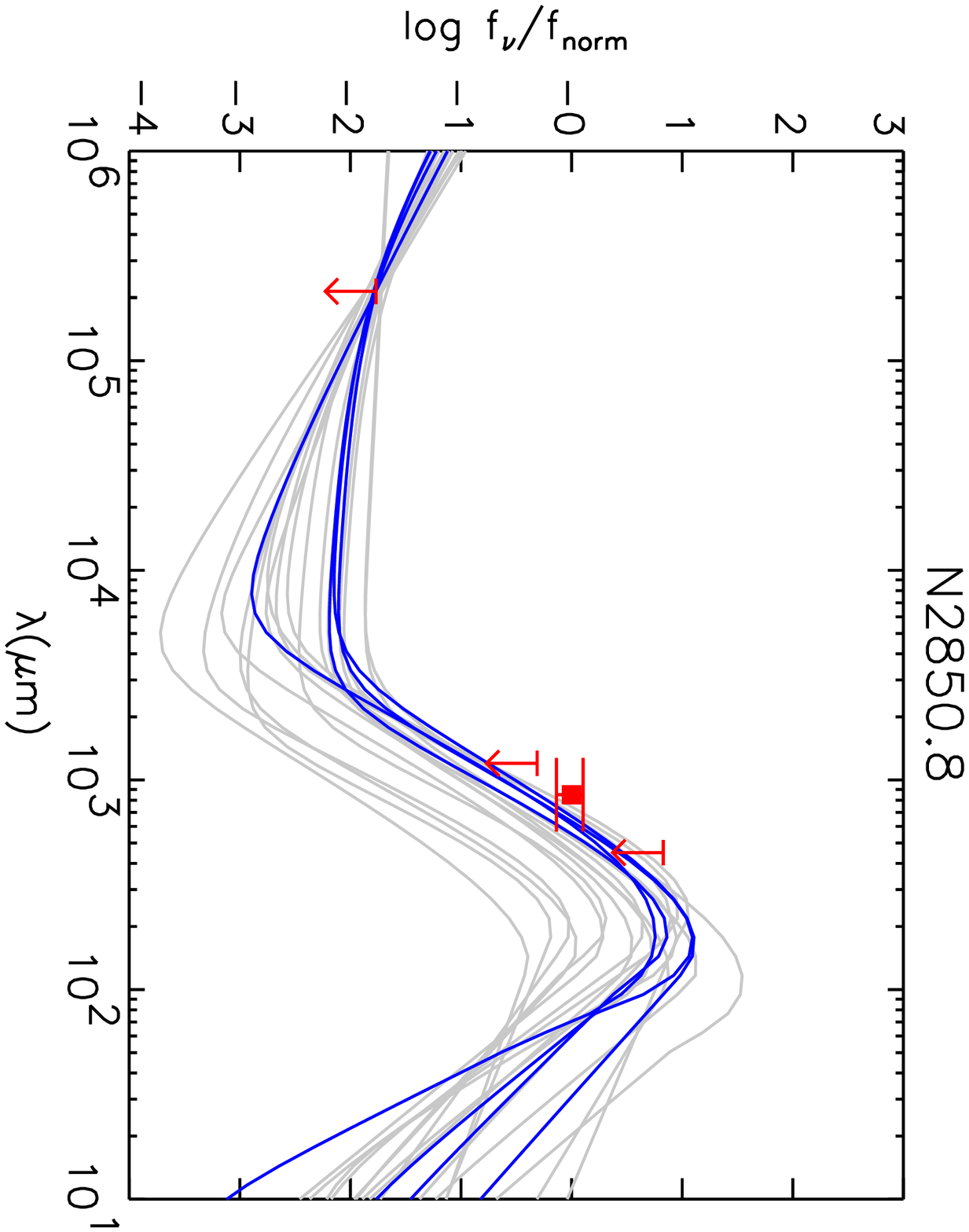}{90} }

    \vspace*{-1.7cm}
    \hbox{ \hspace{0.cm} \figl{4.5cm}{26}{26}{573}{737}{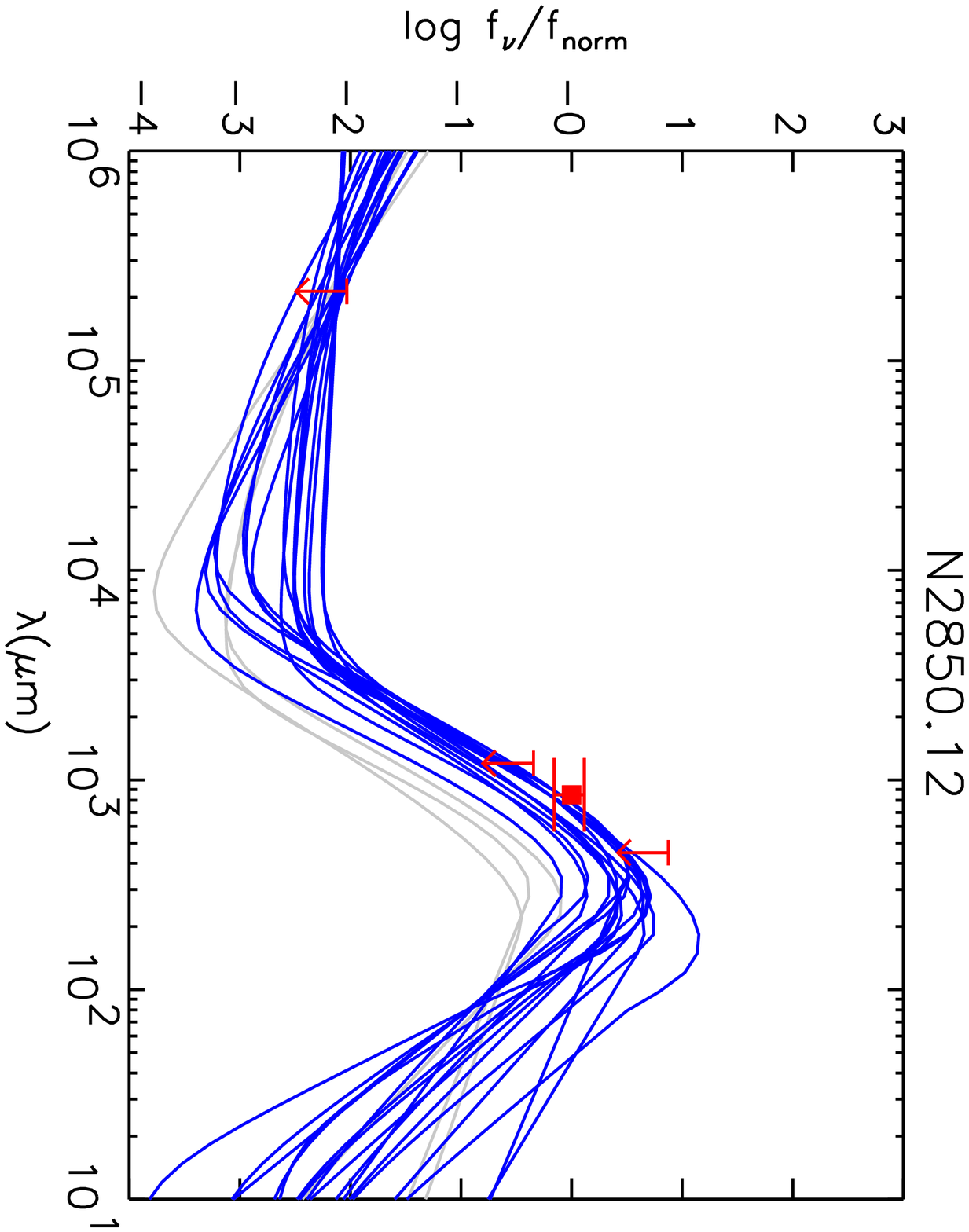}{90}
           \hspace{-13cm} \figl{4.5cm}{26}{26}{573}{737}{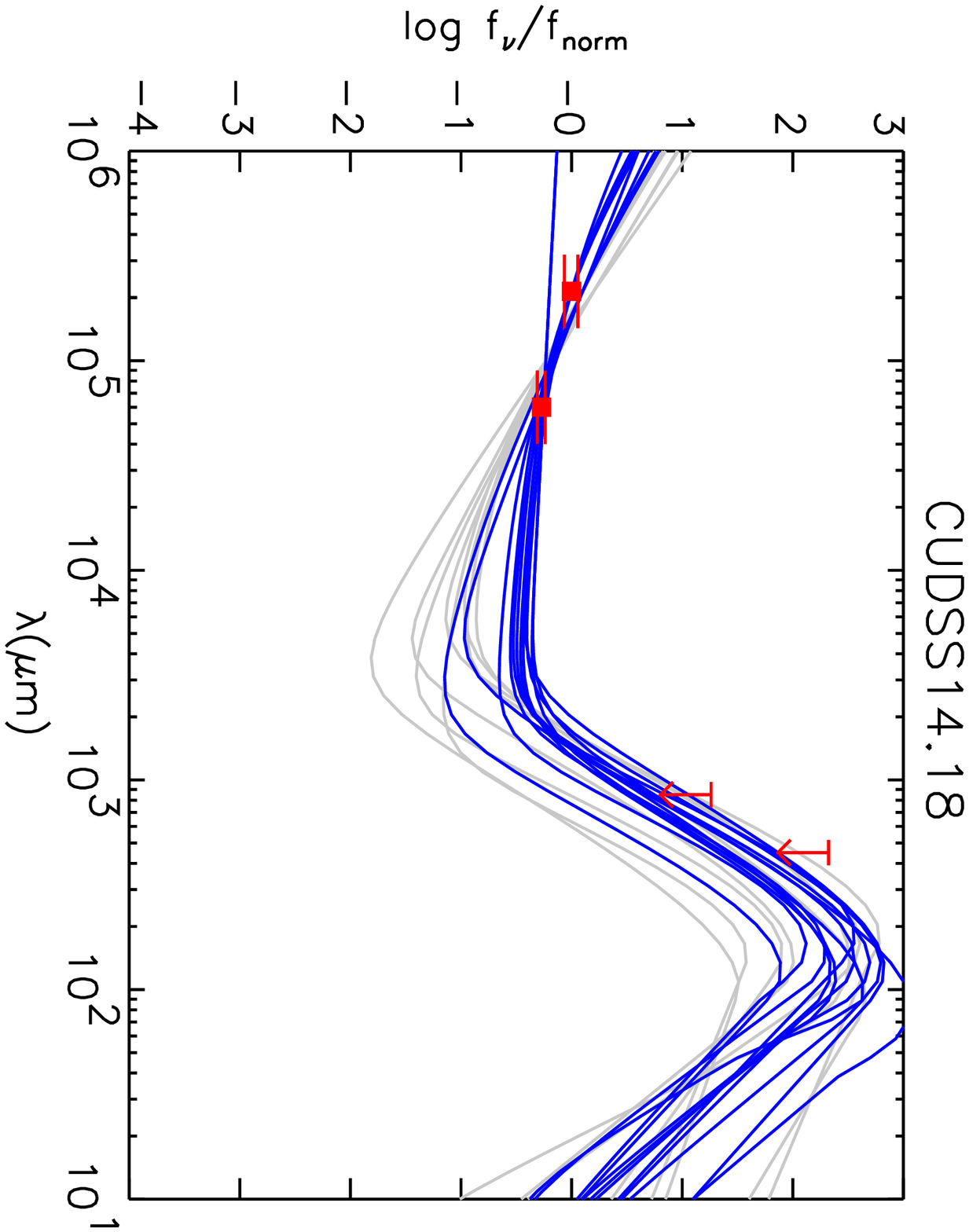}{90}
          \hspace{-13cm} \figl{4.5cm}{26}{26}{573}{737}{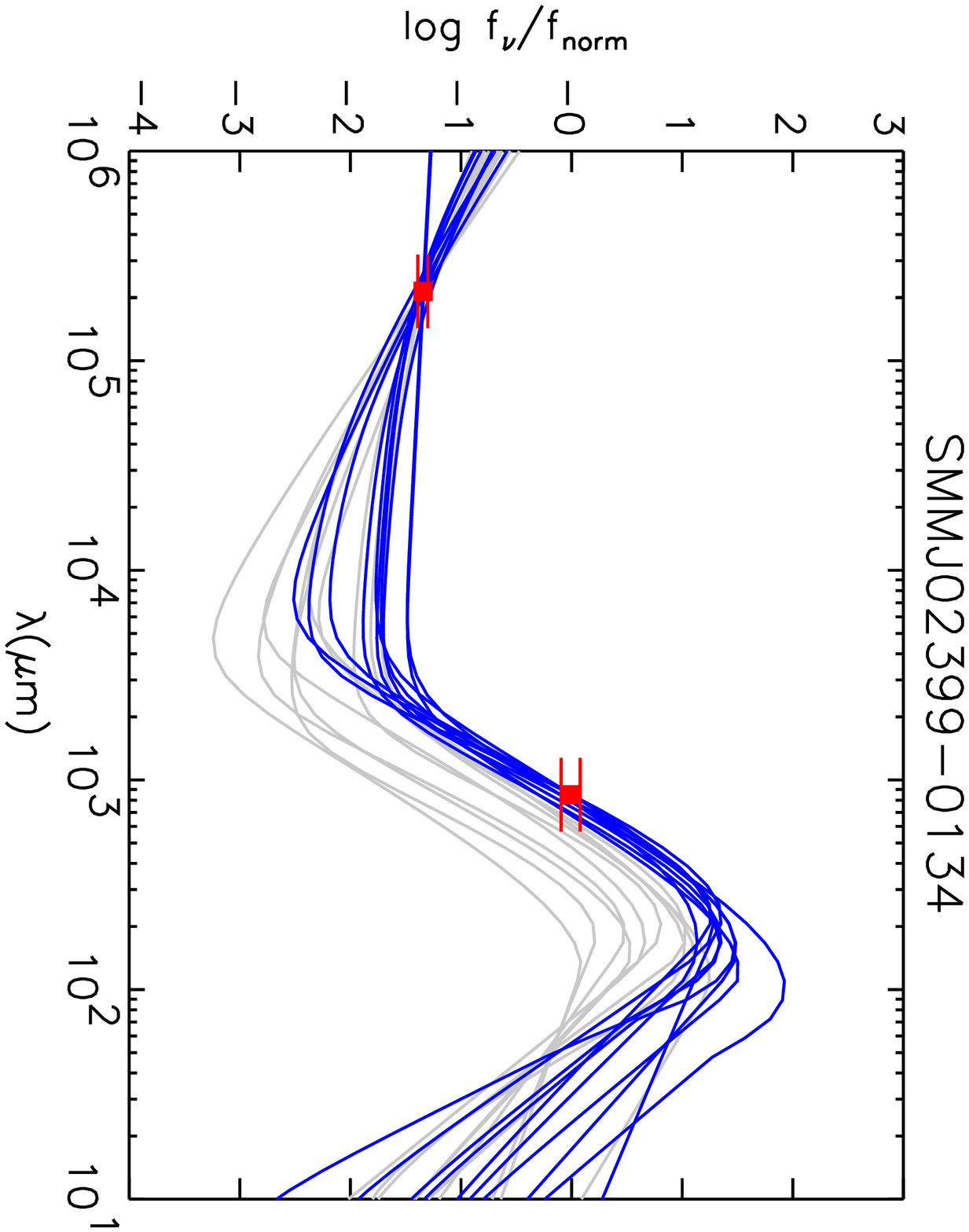}{90} }

    \vspace*{-1.7cm}
    \hbox{ \hspace{0.cm} \figl{4.5cm}{26}{26}{573}{737}{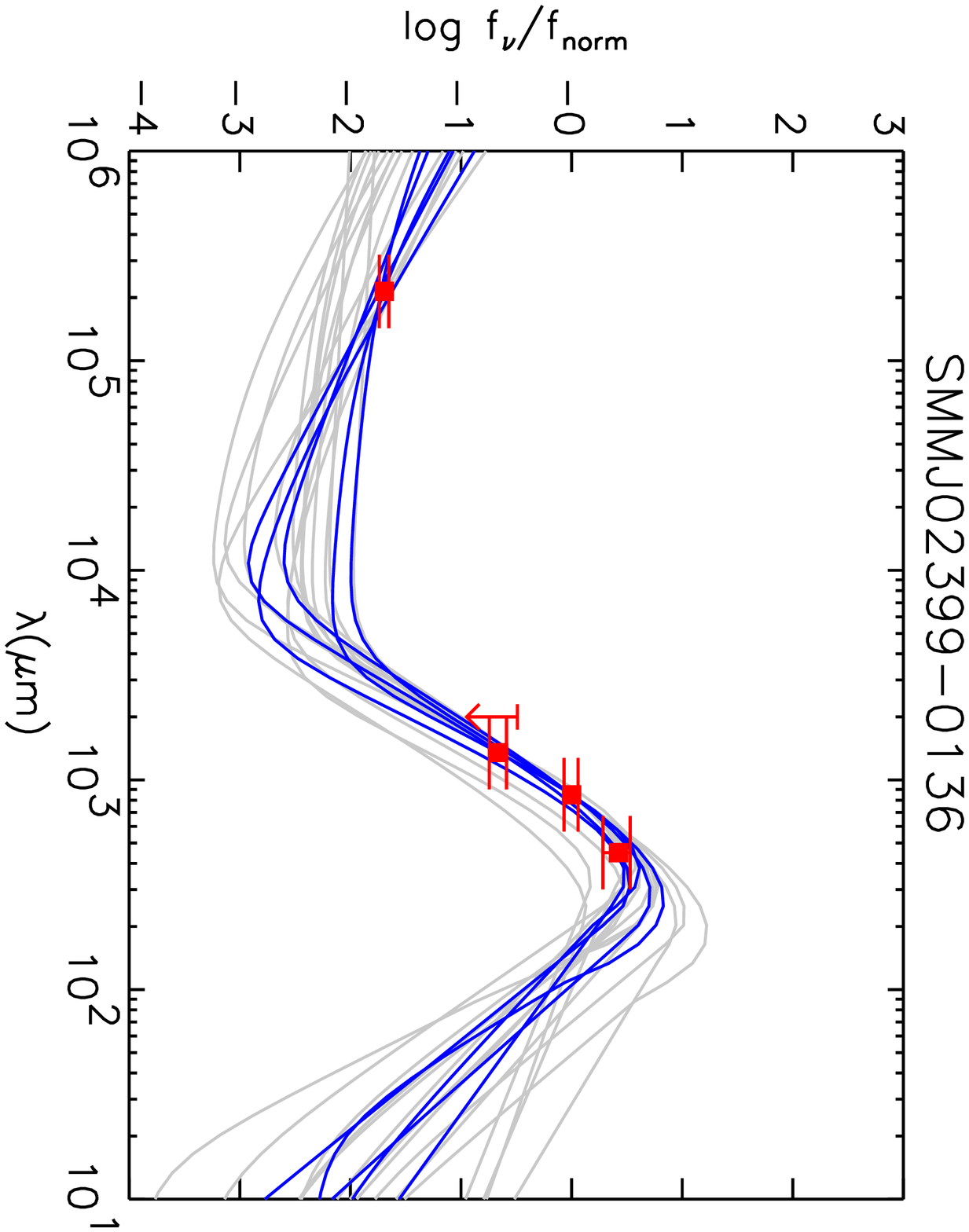}{90}
           \hspace{-13cm} \figl{4.5cm}{26}{26}{573}{737}{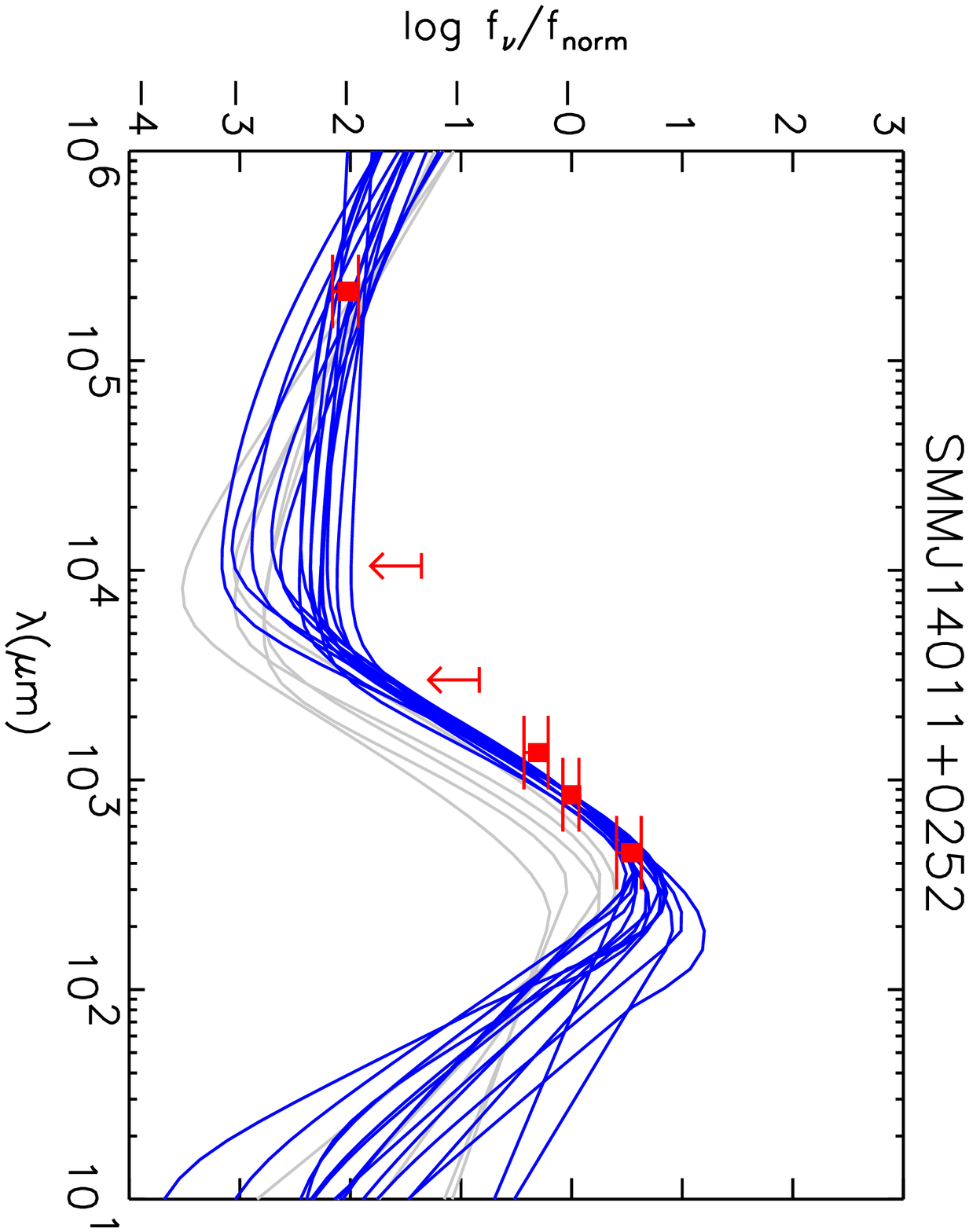}{90}
          \hspace{-13cm} \figl{4.5cm}{26}{26}{573}{737}{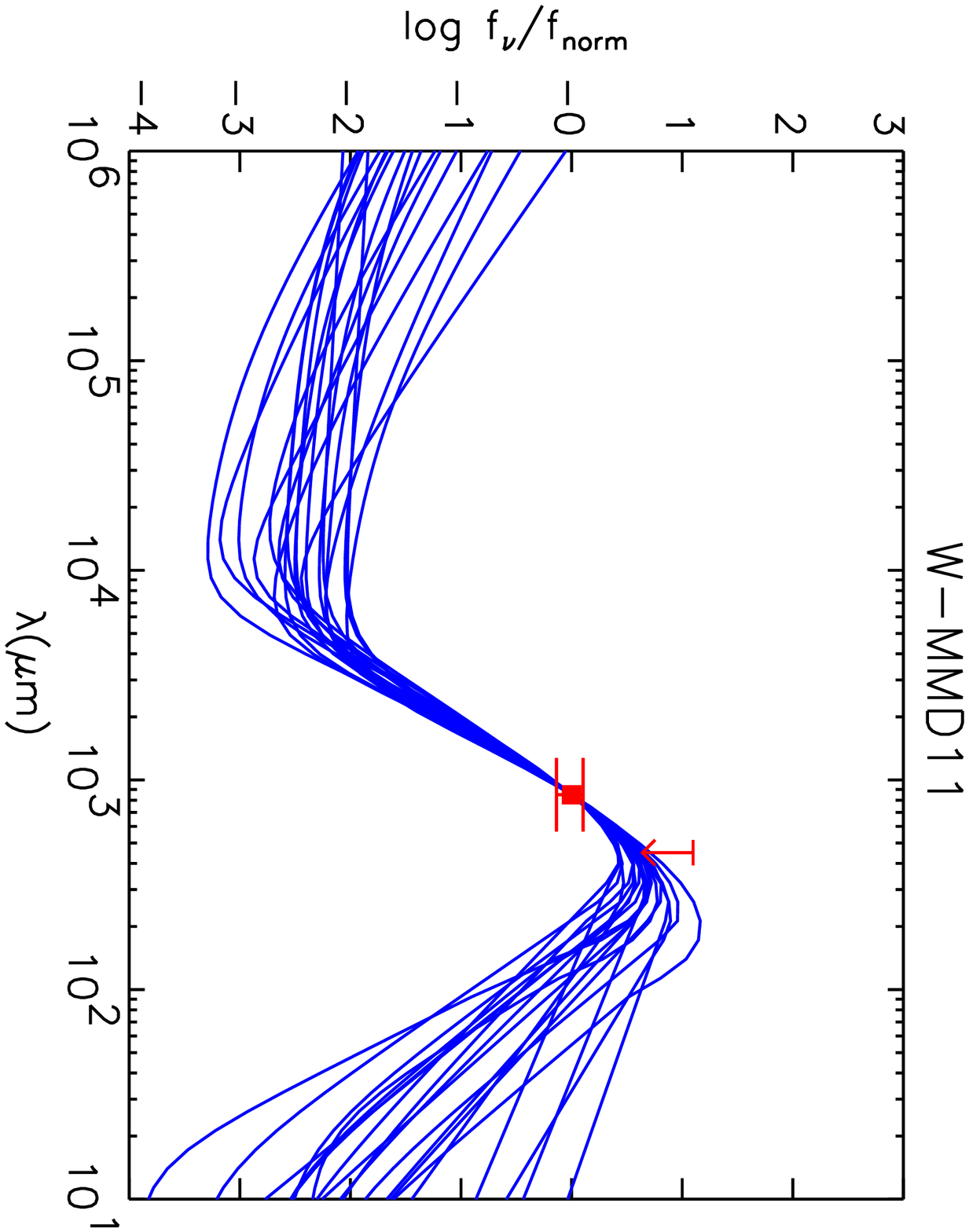}{90} }
    \vspace*{-1.7cm}
    \hbox{ \hspace{0.cm} \figl{4.5cm}{26}{26}{573}{737}{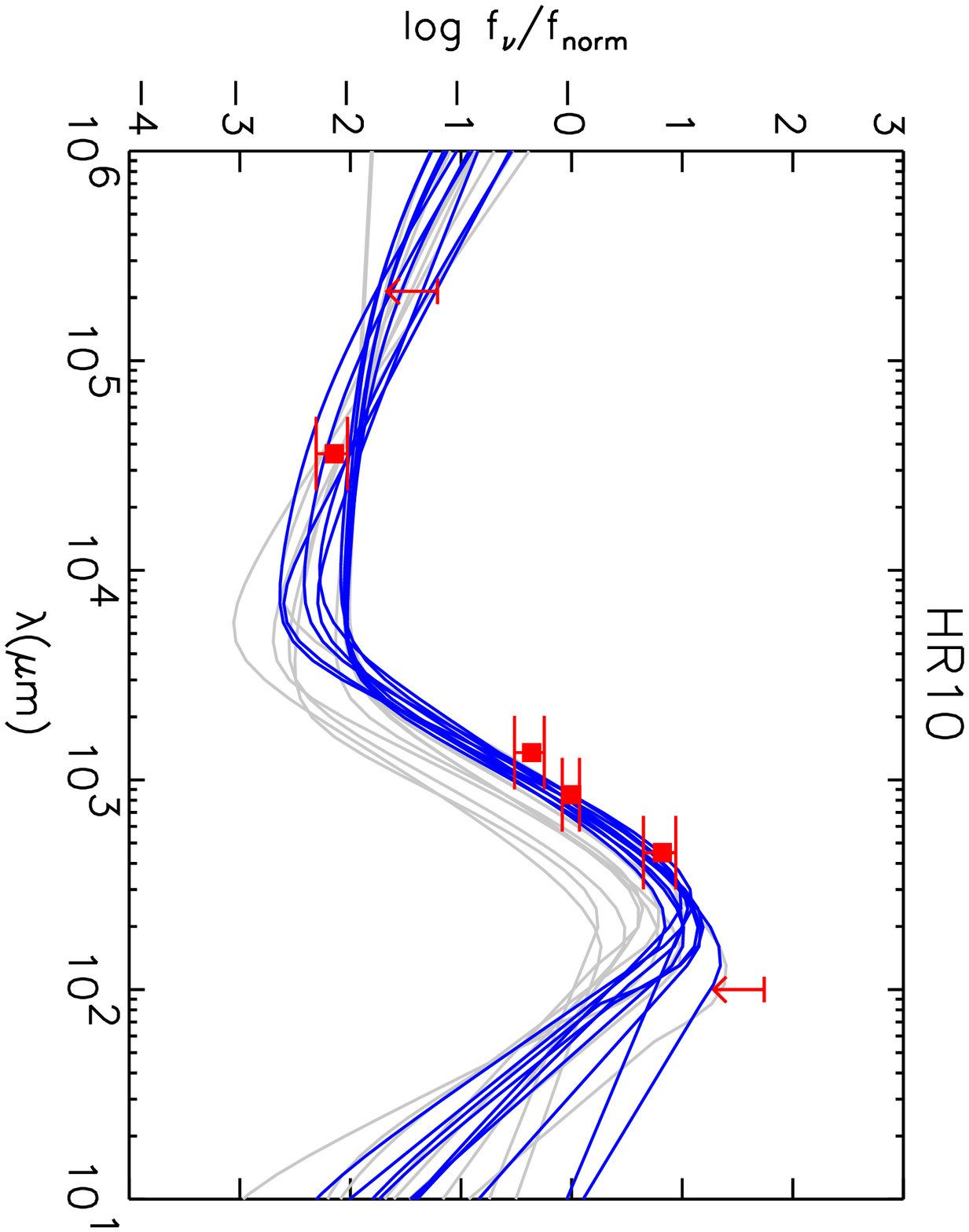}{90}
           \hspace{-13cm} \figl{4.5cm}{26}{26}{573}{737}{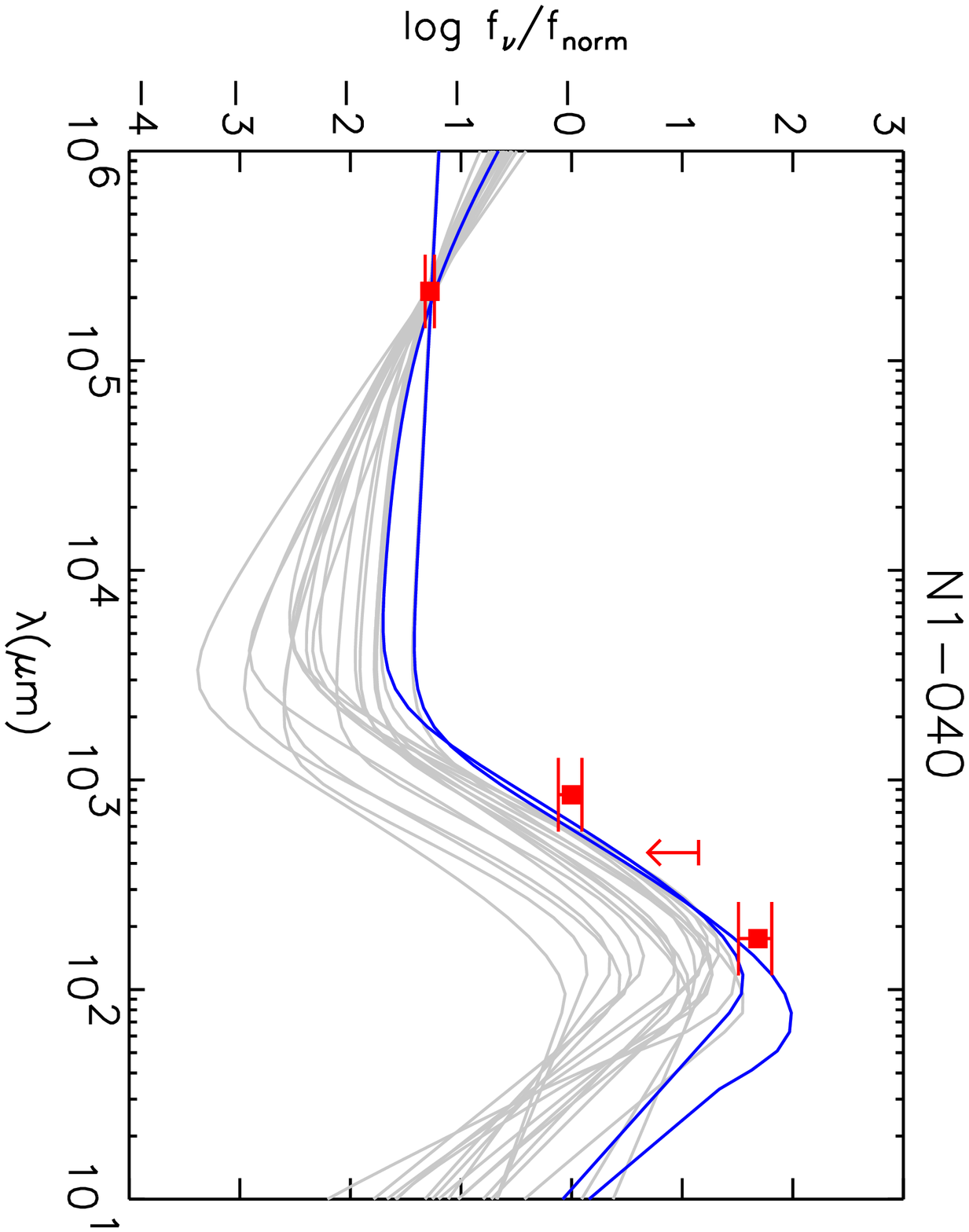}{90}
          \hspace{-13cm} \figl{4.5cm}{26}{26}{573}{737}{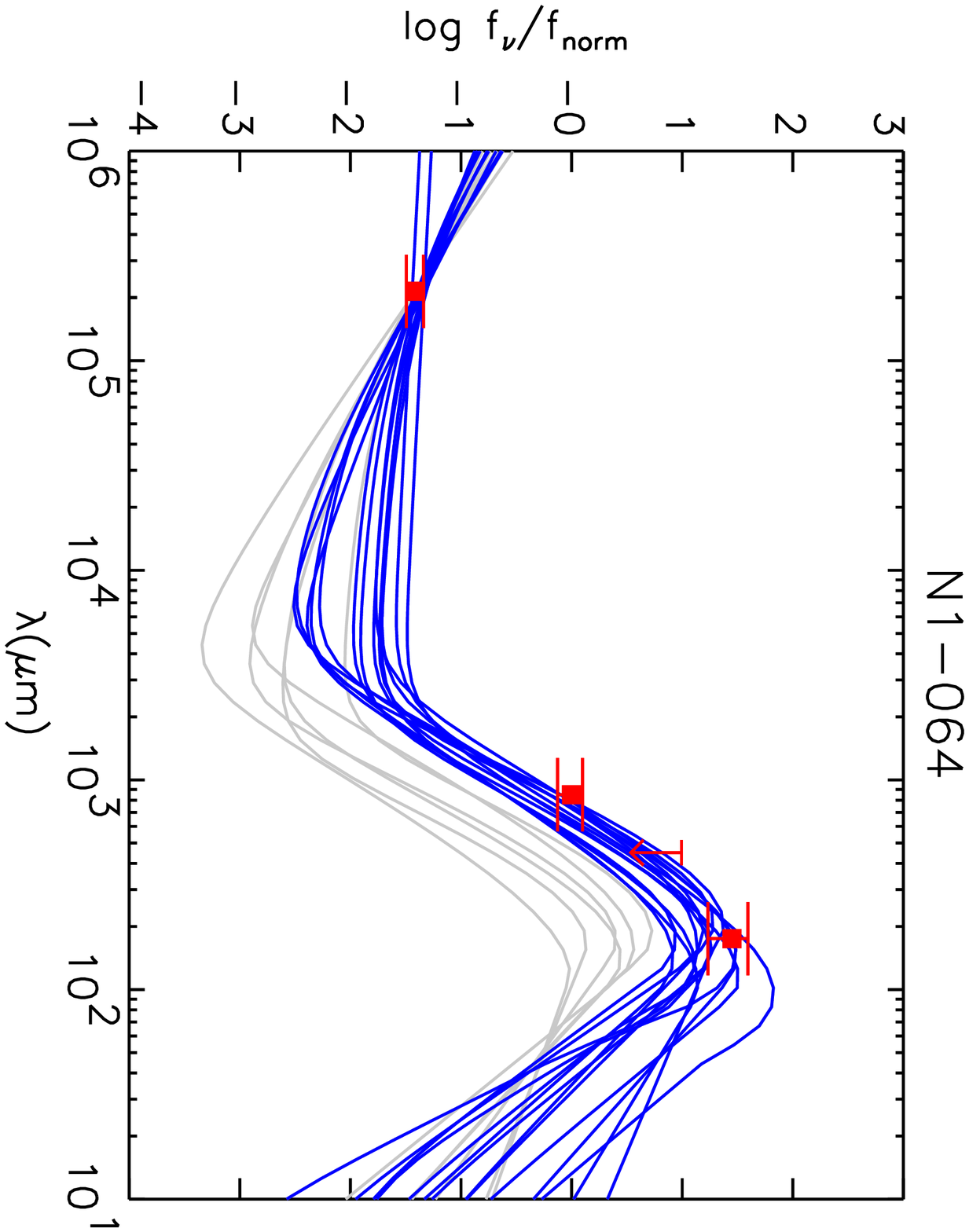}{90} }

\caption{Observed SEDs of sources for which acceptable photometric
redshifts were derived. The SEDs, normalised to the flux density at 
850$\mu$m are shown as squares and arrows.  The arrows indicate 3$\sigma$ upper
limits. The squares denote detection at a level \hbox{$\geq 3\sigma$},
with 1$\sigma$ error bars.  The template SEDs (lines) are redshifted
to the spectroscopic redshift published in the literature, and scaled
to maximize the likelihood of detections and upper limits through
survival analysis (Isobe, Feigelson \& Nelson 1986).  The template SEDs
at this redshift compatible within $3\sigma$ with the observations of the
sources are displayed as darker lines. }
\end{figure*}

\begin{figure}
\vspace*{-1.5cm}
\hspace*{5cm}\figl{5cm}{126}{126}{573}{737}{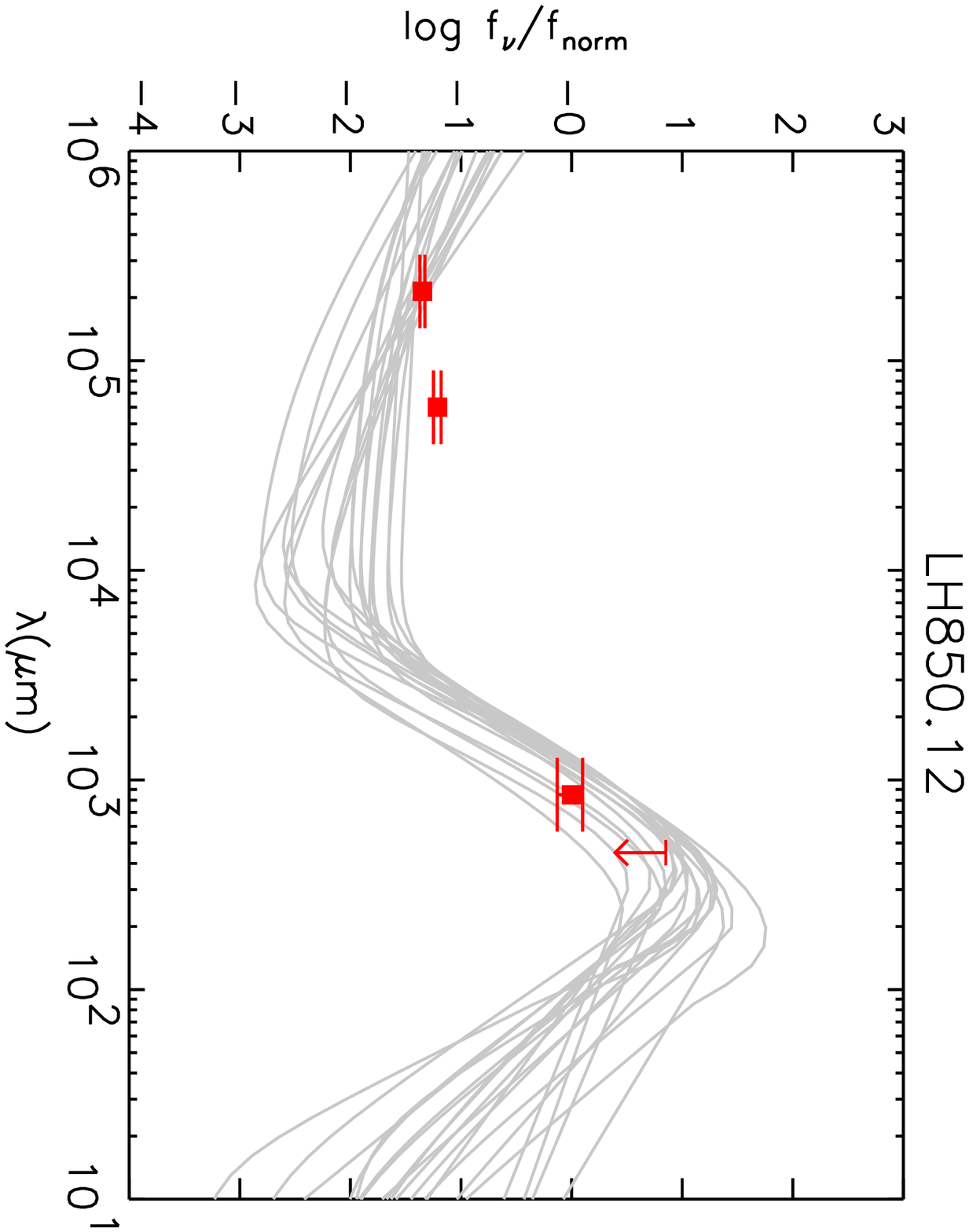}{90}
\vspace*{1.0cm}
\hspace*{5cm}\figl{5cm}{126}{126}{573}{637}{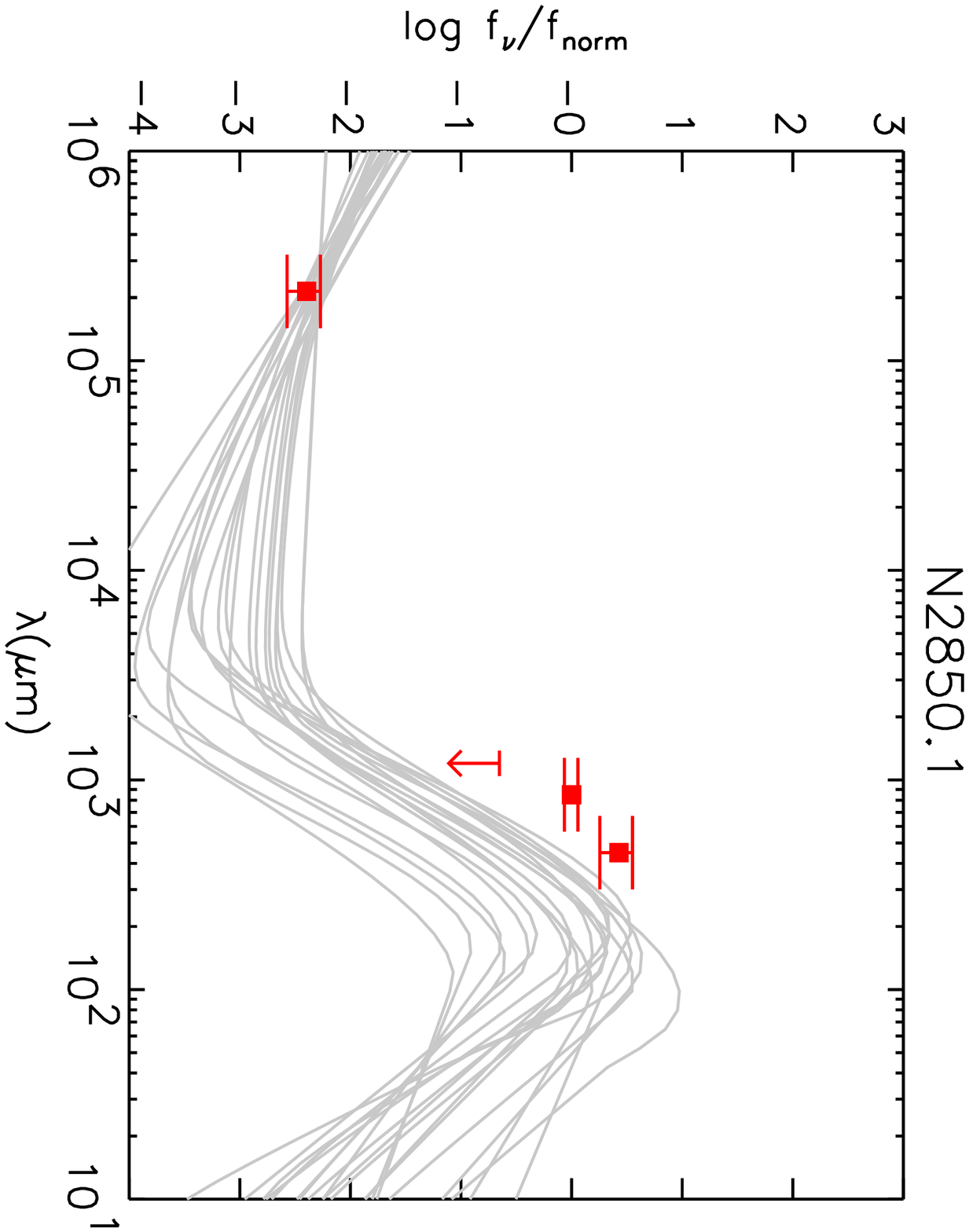}{90}
\caption{Observed SEDs of 2 sub-mm galaxies for which catastrophic
photometric redshifts were derived.  Symbols and lines are as in
Fig.~2.  LE850.12 has been excluded from the photometric-redshift
analysis due to its high and variable radio flux (\S 2.1), but we show
it here to demonstrate explicitly that it cannot be reproduced by any of
our template SEDs.  In the case of N2850.1 the SEDs are fitted under
the assumption that the sub-mm galaxy is at $z_{opt}=0.84$ (Chapman et
al. 2002a). A possible explanation (\S 2.2) for this catastrophic
photometric-redshift is that N2850.l is at a higher redshift and
lensed by the foreground optical galaxy at $z=0.84$.}
\end{figure}

\section{Discussion}
Blain et al. (2003) have stated that the technique described 
in Papers I and II 
assumes a narrow-range of local SED templates, with a tight
distribution of dust temperatures, and luminosities, to derive the
photometric redshifts. We emphasise again that the range of dust
temperatures in our template library ranges from 25 -- 65\,K, and that
we include 20 local galaxies (starbursts and AGN) with well-measured
SEDs and FIR luminosities spanning the range $9.0 < {\rm log} L_{\rm
FIR}/L_{\odot} < 12.3$.
Furthermore, it is  misleading to suggest that 
a wide range of dust temperatures
(that may be correlated with luminosity) 
should translate directly into a similar redshift uncertainty in our
calculations, since the redshift distributions are dominated by those
SEDs that best fit the radio--mm--FIR data.
Although all SEDs contribute to the redshift distributions at some level, but
with varying degrees of significance, a well-defined peak can be
measured.

The comparison presented here provides reassurance that, by allowing
the variety of local template SEDs to be selected at random, and then
scaled to the required FIR luminosity to populate the evolving
luminosity function, we have offered the photometric-redshift
method a library of galaxies with a sufficiently broad range of
dust-temperatures, SED shapes and levels of star formation and AGN
activity from which to find a solution.  We also note that the
sensitivities of the current submillimetre experiments in blank-field
surveys select only those starburst galaxies with $L_{\rm
FIR}/L_{\odot} > 10^{11}$. This being the case, then perhaps the
future choice of SEDs should be restricted to those galaxies more
luminous than this sensitivity limit, in which case our library of
SEDs will only be lacking the highest-luminosity local counterparts
($12.3 < L_{\rm FIR}/L_{\odot} < 13$). If we consider that there
exists a luminosity--temperature dependence in the SEDs of starbursts,
then we will be missing the SEDs that peak at the shortest wavelengths
in our library, and thus we will be underestimating the redshifts for
some fraction of SCUBA sources.

An encouraging aspect of our study is that the derived errors in our
calculations of the individual photometric redshifts (i.e. the 68\%
confidence levels, shown as dotted lines in Fig.~1 and in column~3 of
Table~1), are systematically larger than the measured dispersion
around the $z_{phot}=z_{spec}$ line. This dispersion is based on the
differences between the modes of the photometric redshift
distributions and the spectroscopic redshifts. For instance, the
average 68\% confidence interval is $\delta z \approx 0.65$ for those
sources with two or more colours derived from three or more detections
(black diamonds in Fig.~1), while the measured accuracy of their
photometric redshifts is $\delta z\approx 0.38$ and $\delta z\approx
0.20$, including and excluding N2850.1.  This suggests that our errors
are overestimated, possibly due to considering a range of SED
templates wider than the true, but still unknown, distribution of
SED shapes of sub-mm galaxies.  We note, however, that, with the currently
available photometry, all the SEDs in our template library are
accepted by more than one sub-mm galaxy as possible counterparts
(Fig.~2).

\begin{figure}
\hspace*{7.0cm} 
\figl{8.0cm}{189}{192}{810}{530}{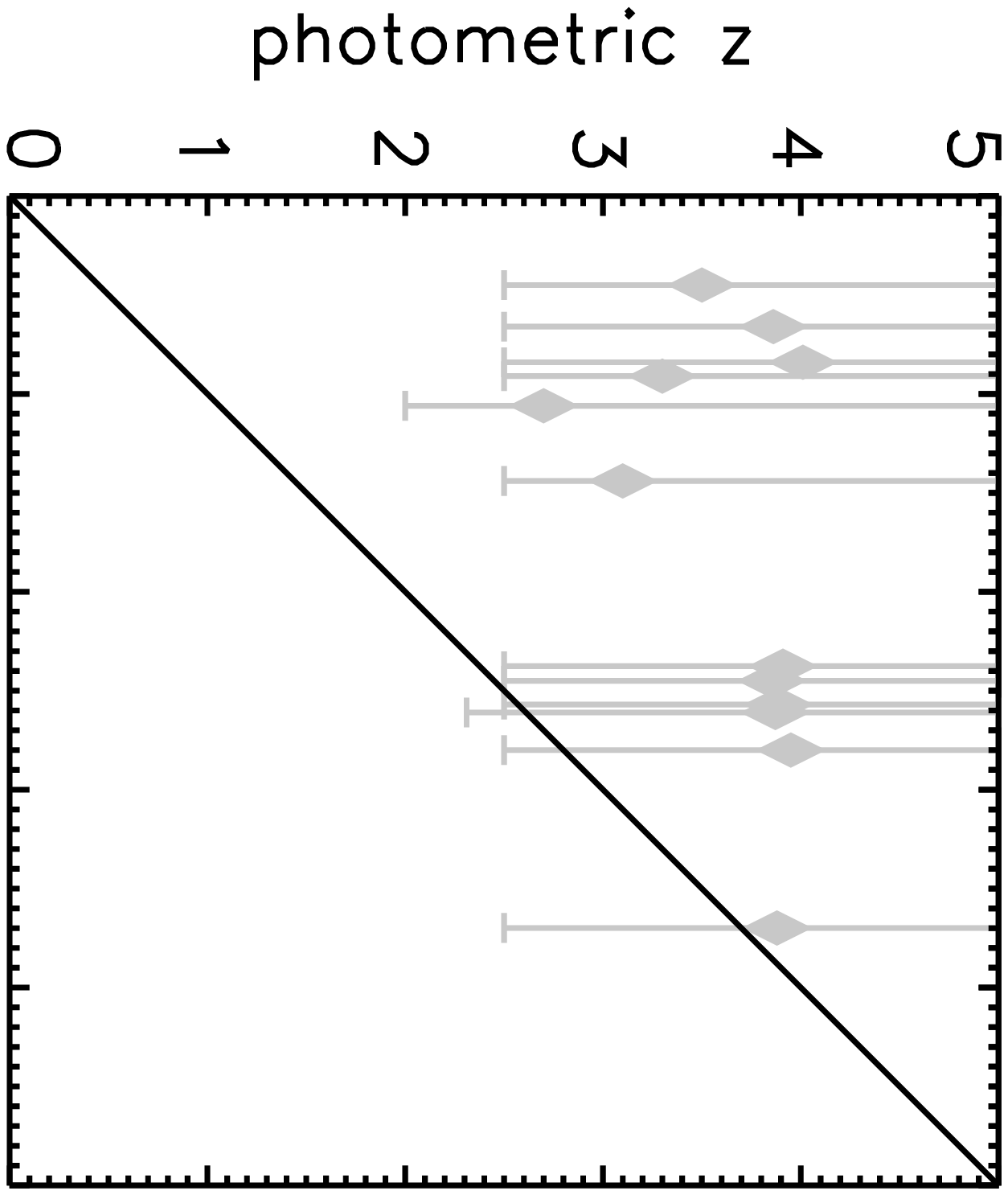}{90}\\
\hspace*{7.0cm} 
\figl{8.0cm}{189}{192}{810}{545}{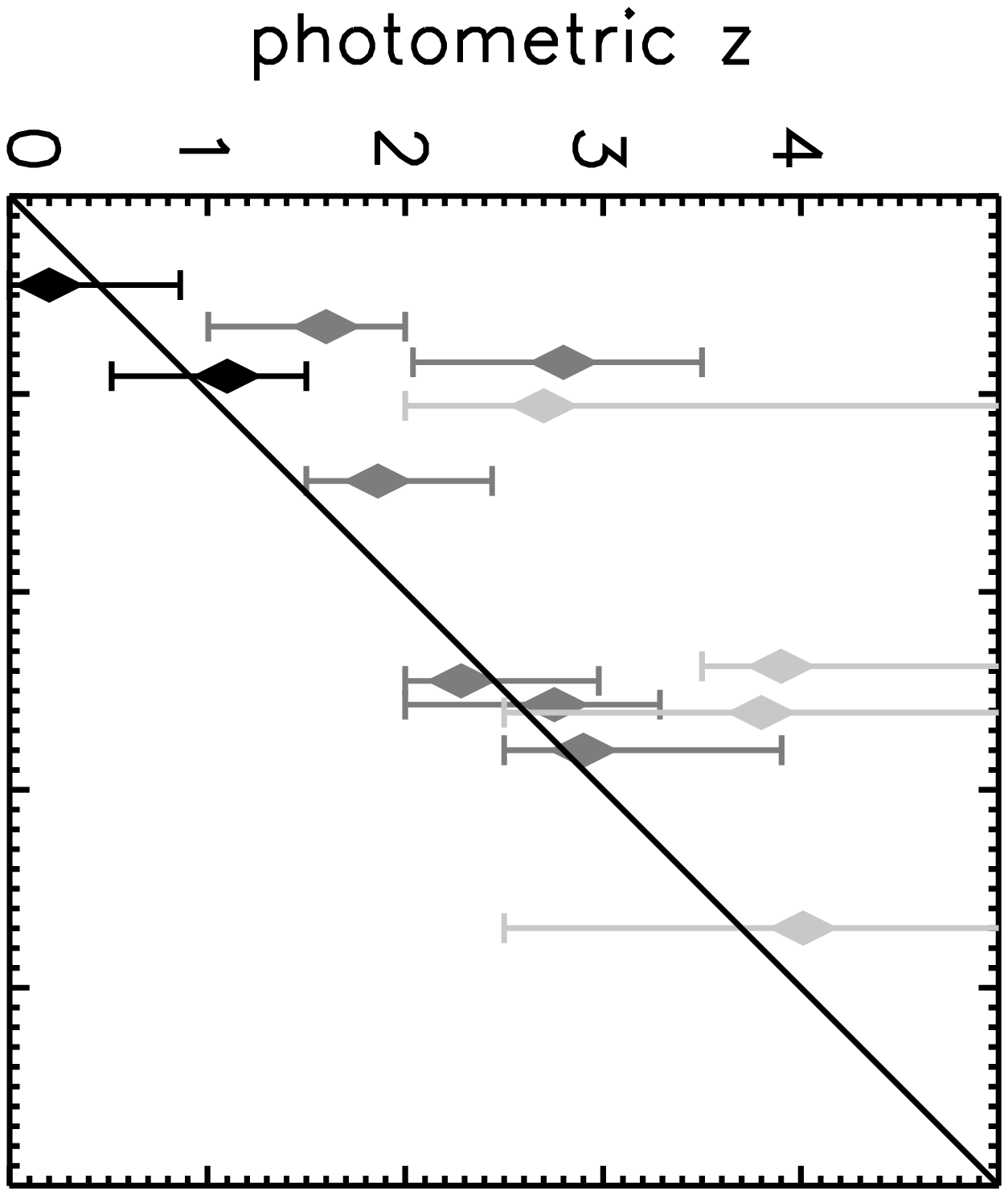}{90}\\
\hspace*{7.0cm} 
\figl{8.0cm}{189}{192}{810}{545}{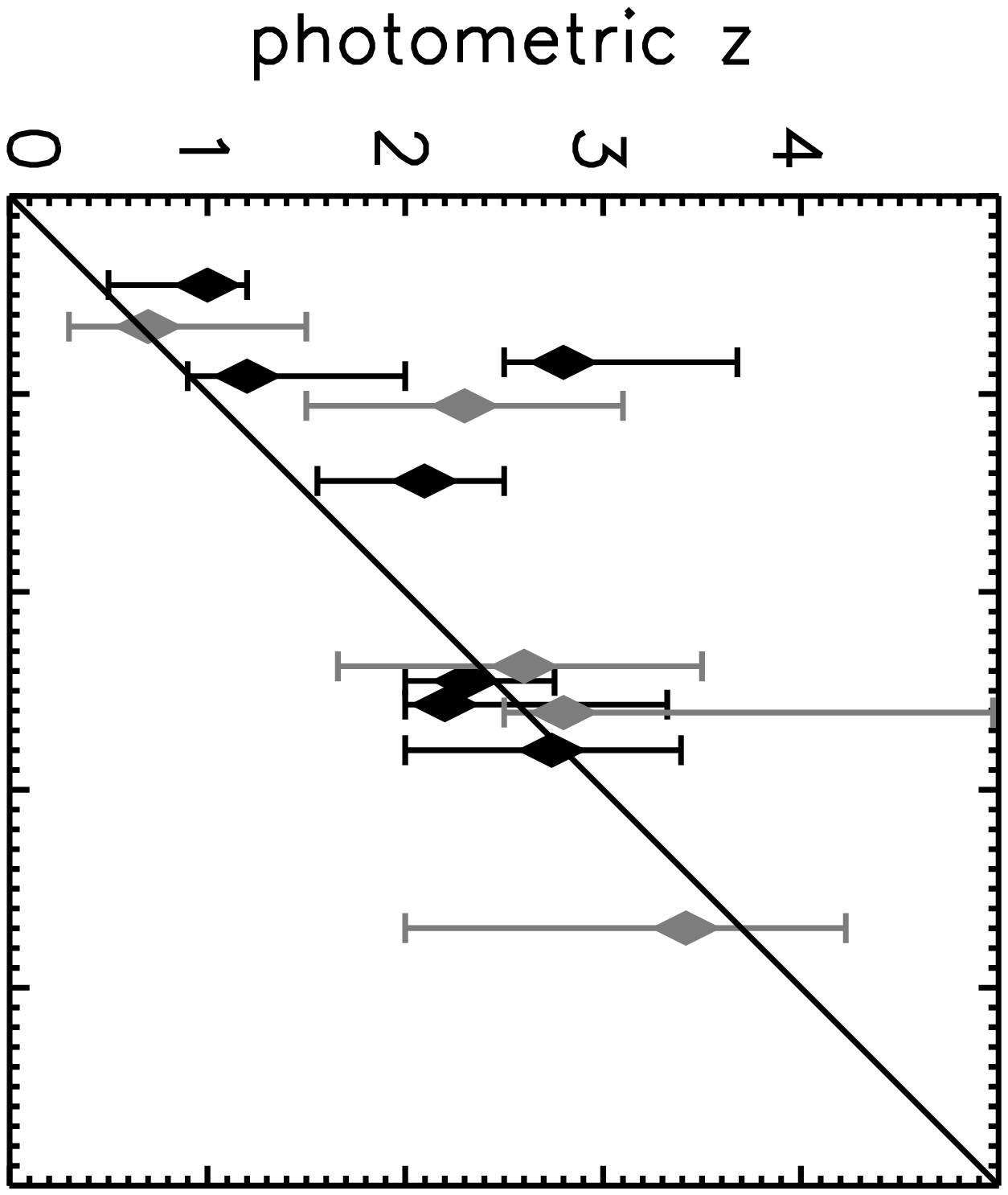}{90}\\
\hspace*{7.0cm} 
\figl{8.0cm}{189}{192}{810}{545}{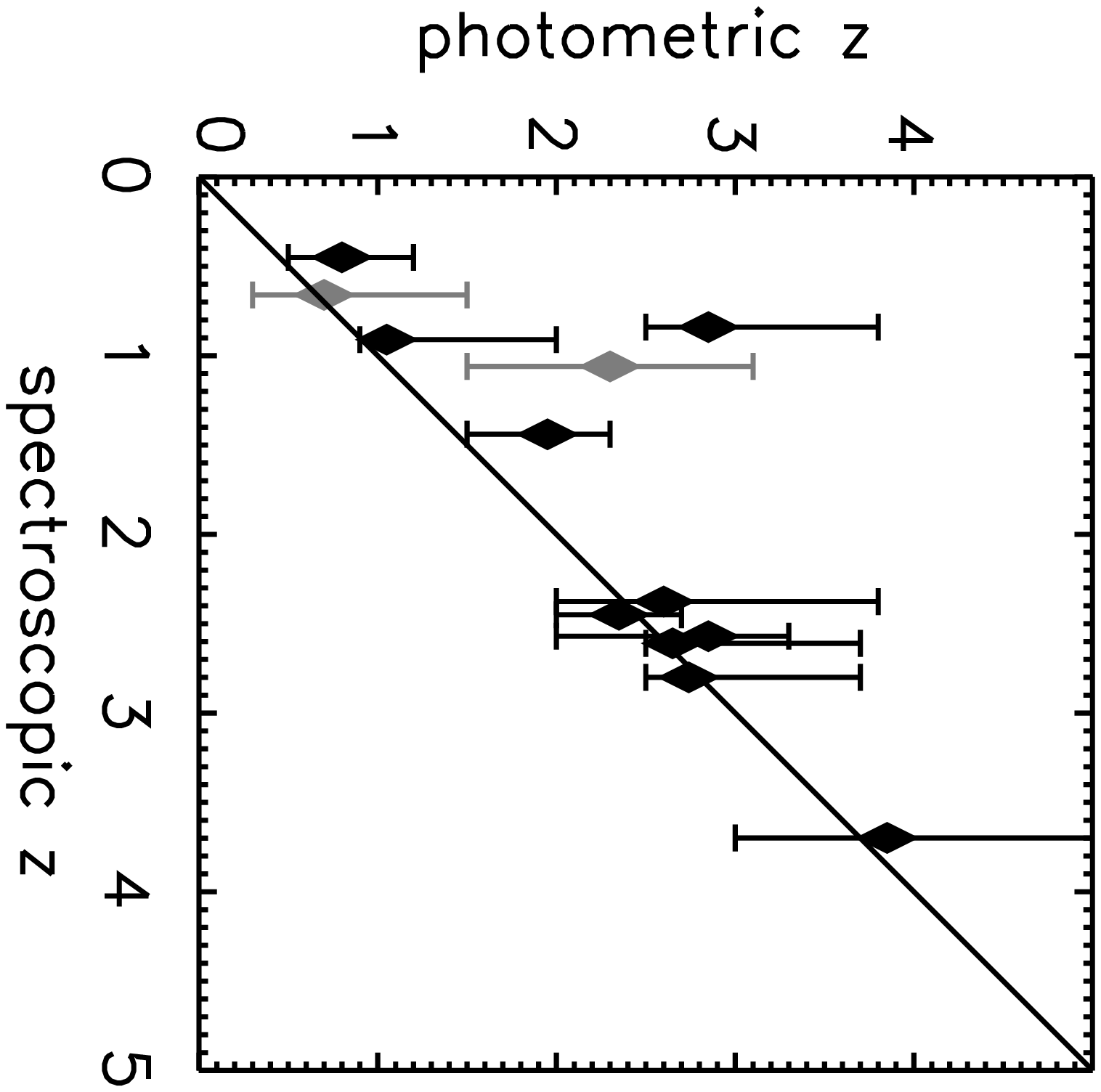}{90}\\
\vspace{1.1cm}
\caption{Dissection of the role played by progressive addition of the 
available photometric data in the final construction of the 
comparison of estimated and spectroscopic redshifts presented in Figure 1.
From top to bottom, the photometric redshifts have been calculated 
using (a) only
the 850$\mu$m flux, (b) the 850 and 450$\mu$m fluxes and upper limits,
and if available (for N-40, N64 and HR10) also the ISO 170$\mu$m and
IRAS 100$\mu$m fluxes and upper limits; (c) 850, 450, 170, 100$\mu$m
fluxes combined with radio data (1.4 and 8GHz fluxes and upper limits); (d)
all available data, including 1--3mm fluxes and upper limits. The
colour code indicates sources with just one robust flux determination
($\ge 3\sigma$) in light grey, sources with at least two robustly
determined fluxes in dark grey, and sources with 3 or more robustly
determined fluxes in black.}
\label{evol_photz}
\end{figure}

To conclude this paper, and to provide further justification for the
results presented in \S 2.2, we illustrate in Fig.\ref{evol_photz} how
the accuracy of the photometric-redshift method evolves and improves
as we include an increasing number of robust photometric measurements.
When based on a single detection at 850$\mu$m (Fig.4a) the initial
photometric-redshift estimates reflect only the prior assumption for
the luminosity evolution of the sub-mm population.  The addition of
detections or deep limits at $\lambda \le 450\mu$m, that sample the
rest-frame FIR--sub-mm peak, place initial constraints on the
photometric redshifts (Fig.4b).  The 850$\mu$m--radio spectral index
has been shown to also provide a useful one-colour measure of
photometric-redshifts for the sub-mm population (e.g. Carilli \& Yun
2000), with the ability to robustly identify galaxies at $z>2$ from
those at lower-redshifts.  Hence the use of data in the radio regime
(1.4--8 GHz) in this analysis continues to improve the
photometric-redshift accuracy (Fig.4c) by providing confirmation, and
therefore adding more weight to the results determined from the data
at $\lambda \le 850\mu$m. Finally, with the addition of further
mm-wavelength data at 1--2\,mm, we eventually derive a
photometric-redshift accuracy of $\pm 0.42$ or $\pm 0.28$ for all 12
galaxies with at least one well-measured colour if we include or
reject N2850.1. respectively (Fig.4d).

\section{Conclusions}
We have complemented our previous comparison of photometric-redshifts
and spectroscopic redshifts for 8 submillimetre galaxies (Aretxaga et
al. 2003) with 7 new spectroscopic redshifts (Chapman et al. 2003a,
Simpson et al. 2004) and recently published MAMBO 1.2mm data (Greve 
et al. 2004). 
The increased number of available spectroscopic redshifts for
submillimetre galaxies with sufficient accompanying rest-frame radio
to FIR photometry confirms the reliability of our previous simulations
and predictions for the redshifts of sub-mm sources. This accuracy is
all the more impressive given that the photometric-redshifts for 6 of
new sources were effectively predicted before the additional 7
spectroscopic redshifts were available (Paper II).

If we consider all available data for those sub-mm galaxies that have
at least one colour determination based on two detections, we conclude
that the rest-frame radio-FIR photometric method can provide redshifts
with an accuracy of $\sim \pm 0.28 $ over the redshift interval $ 0.5 < z
< 4$. This redshift uncertainty increases to $\pm 0.42$ if the true
spectroscopic-redshift of the brightest sub-mm source N2850.1 in the
northern ELAIS\,2 field is $z_{opt}=0.840$ (Chapman et
al. 2003a). There is, however, a more natural explanation for the
discrepancy between the photometric-redshift and
spectroscopic-redshift of N2850.1, namely that the sub-mm source is
lensed by the foreground optical counterpart (as already suggested by Chapman
et al. 2002a), analogous to the situation for HDF850.1, the brightest
object in the Hubble Deep Field (Dunlop et al. 2004).

The photometric method currently relies on limited data, restricted to
a few low S/N detections at observed radio, millimetre and
submillimetre wavelengths.  No optimization of the current method has
been made, based on any prior knowledge of those local SED templates
that are consistent with the rest-frame data.  Despite the lack of any
assumption in our simulations and analysis about a dependence between
the shape of a local SED and luminosity of the redshifted
submillimetre source, the method clearly works to a useful accuracy
(e.g. van Kampen et al. 2004).  The inclusion of such a dependence and
measurements with greater S/N can only reduce the dispersion of
possible SEDs that are consistent with the observational data, and
hence this will reduce the width of the redshift probability
distributions for individual targets.

One of the  difficulties in this analysis has been to select from the
limited publically-available information which sub-mm galaxies have
secure spectroscopic redshifts derived from unambiguous optical or
IR counterparts. It is therefore encouraging to look forward to the
next few years as the essential radio, (sub)millimetre and FIR
photometric data become available from facilities such as the VLA,
GBT, LMT, APEX, BLAST, Herschel, Spitzer and Astro-F for complete and
substantial samples of sub-mm sources.
Thus we are confident that the combination of these high
S/N multi-wavelength rest-frame radio--FIR data will generate
photometric-redshifts with accuracies of $\Delta z \lsim 0.3$ for the
majority of the individual sub-mm selected galaxies.  Given this, it
will be possible to accurately measure the entire redshift
distribution and star formation history of the high-$z$ population of
heavily-obscured starburst galaxies without having to measure a
spectroscopic redshift for every sub-mm source.

\section*{Acknowledgements} IA and DHH gratefully acknowledge support from  
CONACYT grants 39548-F and 39953-F.


\end{document}